\documentclass[letterpaper,12pt]{article}
\usepackage[english]{babel}
\usepackage{blindtext}
\usepackage[utf8]{inputenc}
\usepackage[labelformat=empty]{caption}
\usepackage{textcomp}
\usepackage{comment}
\usepackage{color, soul}
\usepackage{courier}
\usepackage[super]{cite}
\usepackage[title]{appendix}
\usepackage{enumitem}
\usepackage{booktabs}
\usepackage{makecell}
\usepackage{tabularx}
\usepackage{setspace}

\usepackage{achemso}

\usepackage{verbatim}

\usepackage{listings}
\usepackage{color}
\usepackage[svgnames]{xcolor}

\input{listings-stata.tex}

\definecolor{codegreen}{rgb}{0,0.6,0}
\definecolor{codegray}{rgb}{0.5,0.5,0.5}
\definecolor{codepurple}{rgb}{0.58,0,0.82}
\definecolor{backcolour}{rgb}{0.95,0.95,0.92}

\lstdefinestyle{mystyle}{
    backgroundcolor=\color{backcolour},   
    commentstyle=\color{codegreen},
    keywordstyle=\color{magenta},
    numberstyle=\tiny\color{codegray},
    stringstyle=\color{codepurple},
    basicstyle=\ttfamily\footnotesize,
    breakatwhitespace=false,         
    breaklines=true,                 
    captionpos=b,                    
    keepspaces=true,                 
    numbers=left,                    
    numbersep=5pt,                  
    showspaces=false,                
    showstringspaces=false,
    showtabs=false,                  
    tabsize=2
}

\usepackage{graphicx}
\graphicspath{ {./images/} }

\usepackage{amsmath}
\usepackage[fleqn]{mathtools}

\usepackage{adjustbox}
\usepackage{afterpage}
\usepackage{tikz}
\usetikzlibrary{positioning, calc, shapes.geometric, shapes.multipart, shapes, arrows.meta, arrows, decorations.markings, external, trees}

\tikzstyle{Arrow} = [thick, decoration={markings,mark=at position 1 with {\arrow[thick]{latex}}},shorten >= 3pt, preaction = {decorate}]
\tikzstyle{Arrow2} = [thick, decoration={markings,mark=at position 0.9999 with {\arrow[thick]{latex}}},shorten >= 3pt, preaction = {decorate}]
\tikzstyle{Arrow3} = [thick, decoration={markings,mark=at position 0.9995 with {\arrow[thick]{latex}}},shorten >= 3pt, preaction = {decorate}]

\usepackage{blindtext}

\usepackage [autostyle, english = american]{csquotes}
\MakeOuterQuote{"}

\title{Estimating Sibling Spillover Effects with Unobserved Confounding Using Gain-Scores}
\author{David C. Mallinson$^{1,2}$ \and Felix Elwert$^{2,3}$}
\date{%
	\normalsize{$^1$Department of Population Health Sciences, University of Wisconsin-Madison\\%
	$^2$Center for Demography and Ecology, University of Wisconsin-Madison\\%
	$^3$Department of Sociology, University of Wisconsin-Madison\\[2ex]%
	\today}
	}

\begin{document}

\pagenumbering{gobble}

\maketitle

\small{}
\begin{center}
\textbf{Abstract}
\end{center}

A growing area of research in epidemiology is the identification of health-related sibling spillover effects, or the effect of one individual’s exposure on their sibling’s outcome. The health and health care of family members may be inextricably confounded by unobserved factors, rendering identification of spillover effects within families particularly challenging. We demonstrate a gain-score regression method for identifying exposure-to-outcome spillover effects within sibling pairs in a linear fixed effects framework. The method can identify the exposure-to-outcome spillover effect if only one sibling’s exposure affects the other’s outcome; and it identifies the difference between the spillover effects if both siblings’ exposures affect the others’ outcomes. The method fails in the presence of outcome-to-exposure spillover and outcome-to-outcome spillover. Analytic results and Monte Carlo simulations demonstrate the method and its limitations. To exercise this method, we estimate the spillover effect of a child’s preterm birth on an older sibling’s literacy skills, measured by the Phonological Awareness Literacy Screening-Kindergarten test. We analyze 20,010 sibling pairs from a population-wide, Wisconsin-based (United States) birth cohort. Without covariate adjustment, we estimate that preterm birth modestly decreases an older sibling’s test score (-2.11 points; 95\% confidence interval: -3.82, -0.40 points). In conclusion, gain-scores are a promising strategy for identifying exposure-to-outcome spillovers in sibling pairs while controlling for sibling-invariant unobserved confounding in linear settings. 

\bigskip
\scriptsize{}

\noindent \textbf{Acknowledgements}: This work was supported by the Eunice Kennedy Shriver National Institute for Child Health and Human Development (T32 HD007014-42), the University of Wisconsin-Madison Clinical and Translational Science Award program through the National Institutes of Health National Center for Advancing Translational Sciences (Grant UL1TR00427), the University of Wisconsin-Madison School of Medicine and Public Health’s Wisconsin Partnership Program, and the University of Wisconsin-Madison Institute for Research on Poverty. We thank the Wisconsin Department of Children and Families, Department of Health Services, and Department of Public Instruction for the use of data. We also thank Steven T. Cook, Dan Ross, Jane A. Smith, Kristen Voskuil, and Lynn Wimer for data access and programming assistance. We thank Michael Sobel for methodological discussions and Deborah B. Ehrenthal for feedback on this manuscript. The content is solely the responsibility of the authors and does not necessarily represent the official views of supporting agencies. Supporting agencies do not certify the accuracy of the analyses presented. Conflicts of interest: none.

\small{}

\newpage

\tableofcontents

\newpage

\pagenumbering{arabic}
\setcounter{page}{1}


\section{Introduction}

A sibling spillover effect (i.e., “interference” or “carryover effect”) is the effect of an individual’s exposure on their sibling’s outcome.\cite{ogburn2014, vwbook, sjolander2016} The past two decades of epidemiologic research witnessed a burgeoning interest in the role of family environments in childhood health, calling attention to the importance of spillovers within families.\cite{kuh2003, lawlorbook, liu2010, feinberg2012, viner2015, benshlomo2016, deneve2017, morris2017} Yet, sibling spillovers are largely unexamined in the epidemiologic literature, as most field-specific advancements in spillover identification have been restricted to infectious diseases.\cite{halloran1991, halloran1995, longini1998, hudgens2008, vw2011, clemens2011, halloran2012, tt2012, vw2012, halloran2016, benjaminchun2018} With growing interest on the familial interdependence of health,\cite{lawlorbook, liu2010, feinberg2012, viner2015, benshlomo2016, deneve2017, morris2017}  the need for analytical tools to identify sibling spillovers is apparent.  

Unobserved confounding is particularly salient with sibling spillovers. Siblings often share experiences that cultivate their development, which may be unmeasured even in data-rich contexts.\cite{deneve2017} Fixed effect (FE) designs that control for unobserved time-invariant confounding are immediately appealing,\cite{sjolander2016, gunasekara2014, imai2019} but there is little precedent for their use to identify sibling spillovers. Sjölander et al. (2016) investigated FE models with sibling pairs for identifying targeted effects of one siblings’ exposure on their own outcome in the presence of spillover, noting that spillover may be identifiable if only one sibling’s exposure affects the other’s outcome.\cite{sjolander2016} Black et al. (2020) employed a difference-in-differences model with three-sibling clusters for identifying a lower-bound estimate of a child’s disability on an older sibling’s academic performance.\cite{black2020}

In this paper, we demonstrate a method for the identification of one- and two-sided exposure-to-outcome spillovers in sibling pairs with gain-scores (i.e., difference-in-differences, or difference scores), a staple of FE estimation that removes shared confounding by differencing outcomes.\cite{kim2019} We evaluate the gain-score estimator in identifying spillover effects across various models with one-sided or two-sided spillovers. Consistent with the applied FE literature, we focus on linear models with homogenous effects.\cite{gunasekara2014, imai2019, kim2019}

This paper is organized as follows. First, we briefly introduce causal directed acyclic graphs (DAGs), which illustrate our models. Second, we discuss various two-sibling models with one-or two-sided spillover and explain how and when gain-score methods can identify spillover effects. Third, we illustrate our results with simulations. Fourth, we apply the method to identify the effect of a younger sibling's preterm birth on an older sibling's literacy test performance. 


\section{Causal Directed Acyclic Graphs}

\textit{Causal DAGs} are useful for explaining the identification of causal effects. We review necessary terminology for this exposition. Causal DAGs are diagrams consisting of \textit{nodes} (variables) and \textit{directed edges} (direct causal effects) that represent the assumed data-generating process (causal model).\cite{greenland1999, pearlbook, shpitser2012, elwert2013, pearl2013, morganbook} \textit{Paths} are sequences of adjacent edges, regardless of the arrows’ directions. On causal paths between exposure and outcome, all arrows point from the exposure to the outcome. On non-\textit{causal paths} between an exposure and an outcome, at least one arrow points away from the outcome. Causal paths “transmit” causal effects, whereas \textit{non-causal paths} may transmit spurious associations. \textit{Colliders} are variables that receive two inbound arrows on a path (a given variable may be a collider on one path but not on another). Pearl’s \textit{d-separation} criterion determines which variables in data generated by the assumed DAG are conditionally or unconditionally independent: two variables are independent if all paths between them are closed; and a path is closed if it includes a non-collider as intermediate variable that is conditioned on, or if it includes a collider as intermediate variable that is not conditioned on.\cite{pearlbook, shpitser2012, morganbook} Conversely, two variables may be associated if at least one path between them is open (\textit{d-connected}); and a path is open if it is not closed.  

Typically, health researchers attempt to identify causal effects by adjusting for observed variables via regression analysis, matching, or inverse-probability weighting, so that all causal paths between exposure and outcome are open and all non-causal paths between them are closed.\cite{pearlbook, shpitser2012, morganbook}  However, researchers may worry about open non-causal paths with unobserved confounders that cannot be closed by covariate adjustment. In multilevel analyses in which observations are clustered into groups (e.g., children in sibling pairs), FE methods can sometimes identify causal effects by subtracting out certain types of group-level unobserved confounding.\cite{gunasekara2014, imai2019, kim2019} Next, we describe several sibling spillover models with unobserved confounding and show when gain-score estimation—a FE approach—can identify the spillover effect.


\section{Method For Sibling Spillover Identification}

\subsection*{Model and Assumptions}

We first present our baseline sibling spillover model and subsequently introduce variations on this model. For illustration, we discuss the spillover effect of a child’s early health shock (e.g., serious illness) on their sibling’s later academic achievement (e.g., test scores). This example is purposefully generic but broadly applicable, and it draws upon prior work of health-related spillover effects on academic performance\cite{black2020} while motivating our empirical application.

Our baseline model is a linear two-sibling comparison design with one-sided spillover (\textbf{Figure 1A}). Subscript $i=1,\ldots,N$ indicates cluster (family) and subscript $j=1,2$ indicates sibling. $T_{ij}$ represents a binary or continuous exposure (e.g., the health shock), $Y_{ij}$ represents a continuous outcome (e.g., academic performance), $U_i$ represents unobserved family-level confounding (i.e., the FE), and $D_i$ represents the gain-score, $D_i=Y_{i2}-Y_{i1}$. Causal effects in this model include the spillover effect, $\theta$ ($T_{i1}\rightarrow Y_{i2}$), of sibling 1’s exposure on sibling 2’s outcome; the targeted effects, $\delta$ ($T_{ij}\rightarrow Y_{ij}$), of each sibling’s exposure on their own outcome; and confounding effects of the unobserved family-level confounders on each sibling’s exposure and outcome, $\psi\chi$ ($T_{i1}\leftarrow U_i\rightarrow Y_{i1}$)  and $\psi\gamma$ ($T_{i2}\leftarrow U_i\rightarrow Y_{i2}$). 

\afterpage{%
\begin{figure}
\centering
\resizebox{\linewidth}{!}{
\begin{tikzpicture}



\node (1) [scale=1.3] {U$_{i}$};

\node [above right =0.7cm and 1cm of 1, scale=1.3] (2) {T$_{i1}$};
\node [right =3cm of 2, scale=1.3] (3) {Y$_{i1}$};

\node [below right =0.7cm and 1cm of 1, scale=1.3] (4) {T$_{i2}$};
\node [right =3cm of 4, scale=1.3] (5) {Y$_{i2}$};

\node [right =2cm of 5, scale=1.3] (8) {D$_{i}$};

\node [above left =2.8cm and 0.2cm of 1, scale=1.3] (991) {\textbf{(1A)}};


\draw[Arrow] (1.east) -- (2.south west) node [midway, above, scale=1.3] {$\chi$};
\draw[Arrow3] (1) to [out=90, in=130] node [pos=0.6, above, scale=1.3] {$\psi$} (3);
\draw[Arrow3] (1) to [out=270, in=230] node [pos=0.6, below, scale=1.3] {$\psi$} (5);
\draw[Arrow] (1.east) -- (4.north west) node [midway, below, scale=1.3] {$\gamma$};

\draw[Arrow] (2.east) -- (3.west) node [midway, above, scale=1.3] {$\delta$};

\draw[Arrow] (2.south east) -- (5.north west)  node [midway, above, scale=1.3] {$\theta$};

\draw[Arrow] (4.east) -- (5.west) node [midway, above, scale=1.3] {$\delta$};

\draw[Arrow] (3.south east) -- (8.north west) node [midway, above, scale=1.3] {\it -1};
\draw[Arrow] (5.east) -- (8.west) node [midway, above, scale=1.3] {\it +1};




\node [right =10cm of 1, scale=1.3]  (21) {U$_{i}$};

\node [above left =2.8cm and 0.2cm of 21, scale=1.3] (993) {\textbf{(1B)}};

\node [above right =0.7cm and 1cm of 21, scale=1.3] (22) {T$_{i1}$};
\node [right =3cm of 22, scale=1.3] (23) {Y$_{i1}$};

\node [below right =0.7cm and 1cm of 21, scale=1.3] (24) {T$_{i2}$};
\node [right =3cm of 24, scale=1.3] (25) {Y$_{i2}$};

\node [right =2cm of 25, scale=1.3] (28) {D$_{i}$};


\draw[Arrow] (21.east) -- (22.south west) node [midway, above, scale=1.3] {$\chi$};
\draw[Arrow3] (21) to [out=90, in=130] node [pos=0.6, above, scale=1.3] {$\psi$} (23);
\draw[Arrow3] (21) to [out=270, in=230] node [pos=0.6, below, scale=1.3] {$\psi$} (25);
\draw[Arrow] (21.east) -- (24.north west) node [midway, below, scale=1.3] {$\gamma$};

\draw[Arrow] (22.east) -- (23.west) node [midway, above, scale=1.3] {$\delta$};

\draw[Arrow] (22.south east) -- (25.north west)  node [midway, above, scale=1.3] {$\theta$};

\draw[Arrow] (24.east) -- (25.west) node [midway, above, scale=1.3] {$\delta$};

\draw[Arrow] (23.south east) -- (28.north west) node [midway, above, scale=1.3] {\it -1};
\draw[Arrow] (25.east) -- (28.west) node [midway, above, scale=1.3] {\it +1};

\draw[Arrow] (24.north) -- (22.south) node [midway, left, scale=1.3] {$\tau$};




\node [below right =7cm and 5cm of 1, scale=1.3]  (41) {U$_{i}$};

\node [above left =2.8cm and 0.2cm of 41, scale=1.3] (995) {\textbf{(1C)}};

\node [above right =0.7cm and 1cm of 41, scale=1.3] (42) {T$_{i1}$};
\node [right =3cm of 42, scale=1.3] (43) {Y$_{i1}$};

\node [below right =0.7cm and 1cm of 41, scale=1.3] (44) {T$_{i2}$};
\node [right =3cm of 44, scale=1.3] (45) {Y$_{i2}$};

\node [right =2cm of 45, scale=1.3] (48) {D$_{i}$};


\draw[Arrow] (41.east) -- (42.south west) node [midway, above, scale=1.3] {$\chi$};
\draw[Arrow3] (41) to [out=90, in=130] node [pos=0.6, above, scale=1.3] {$\psi$} (43);
\draw[Arrow3] (41) to [out=270, in=230] node [pos=0.6, below, scale=1.3] {$\psi$} (45);
\draw[Arrow] (41.east) -- (44.north west) node [midway, below, scale=1.3] {$\gamma$};

\draw[Arrow] (42.east) -- (43.west) node [midway, above, scale=1.3] {$\delta$};

\draw[Arrow] (42.south east) -- (45.north west)  node [midway, above, scale=1.3] {$\theta$};

\draw[Arrow] (44.east) -- (45.west) node [midway, above, scale=1.3] {$\delta$};

\draw[Arrow] (43.south east) -- (48.north west) node [midway, above, scale=1.3] {\it -1};
\draw[Arrow] (45.east) -- (48.west) node [midway, above, scale=1.3] {\it +1};

\draw[Arrow] (42.south) -- (44.north) node [midway, left, scale=1.3] {$\phi$};

\end{tikzpicture}
}
\caption{\textbf{Figure 1.} Causal directed acyclic graphs for linear data-generating models with one-sided exposure-to-outcome sibling spillover. Subscripts $i$ and $j$ denote cluster and sibling, respectively. $T_{ij}$ is the exposure, $Y_{ij}$ is the outcome, $D_i$ is the gain-score, and $U_i$ is an unobserved family-level confounder. Greek letters denote effects. (1A) does not have exposure-to-exposure spillover, whereas (1B) and (1C) have exposure-to-exposure spillovers.} 
\end{figure}
\clearpage
}

All models embed several simplifying assumptions. First, the targeted effects, $\delta$, of $T_{ij}$ on $Y_{ij}$ and the confounding effects, $\psi$, of $U_i$ on $Y_{ij}$ are equal for both siblings.\cite{gunasekara2014, imai2019} Second, all effects are linear and homogenous. Third, there is only partial interference (i.e., spillovers within sibling clusters but not between sibling clusters).\cite{hudgens2008, sobel2006} Aside from partial interference, these assumptions align with conventional FE models.\cite{sjolander2016, gunasekara2014, imai2019, kim2019} 

Notably, our presentation abstracts from sibling-specific observed baseline covariates, $\textbf{\textit{C}}_{ij}$. Covariates may be added to our baseline and subsequent models as long as one can condition on $\textbf{\textit{C}}_{ij}$ without loss of generality.  

\subsection*{Gain-Score Estimation and Identification of Spillover Effects}

This subsection details the gain-score estimation strategy and demonstrates when our estimator point-identifies spillover effects (i.e., recovers the estimand precisely) for nine sibling spillover models that differ by whether spillover is one- or two-sided and by whether additional spillovers originate from outcomes. 

\subsubsection*{\textit{Gain-score estimation}}

We investigate the ability of a gain-score estimator to identify exposure-to-outcome spillover effects. First, we regress the gain-score on both siblings’ exposures, 

\begin{equation}
D_i=b_1T_{i1}+b_2T_{i2}+e_i,
\end{equation}

\noindent where $b_1$ and $b_2$ are partial regression coefficients for $T_{i1}$ and $T_{i2}$, respectively, and $e_i$ is an error term. We then sum the partial regression coefficients to compute a “spillover coefficient” ($SC$), 

\begin{equation}
SC=b_1+b_2.
\end{equation}

We will now interrogate whether the SC identifies causal spillover effects in each of several commonly assumed data generating processes in health research. 

\subsubsection*{\textit{Settings with one-sided spillover}}

The object of interest (estimand) is $\theta$, or the direct spillover effect of sibling 1’s health shock on sibling 2’s academic performance. Under the baseline model (\textbf{Figure 1A}), three open paths connect $T_{i1}$ and $Y_{i2}$. The first path, $T_{i1}\rightarrow Y_{i2}$, is the causal spillover effect of interest. The other two paths are non-causal paths that may transmit spurious association. The first non-causal path, $T_{i1}\leftarrow U_i\rightarrow T_{i2}\rightarrow Y_{i2}$, can be closed by adjusting on $T_{i2}$. However, the second non-causal path, $T_{i1}\leftarrow U_i\rightarrow Y_{i2}$, cannot be closed by covariate adjustment because it only contains the unobserved variable $U_i$. 

Nonetheless, we can identify $\theta$ through gain-score regression. Under the assumptions of \textbf{Figure 1A}, it can be shown that $b_1=\theta-\delta$ and $b_2=\delta$, using elementary regression algebra. Therefore, the spillover coefficient equals $SC=b_1+b_2=\theta$.

The intuition for this result is that first-differencing exactly offsets confounding biases involving $U_i$,26 and that the $SC$ corrects for the contamination of the spillover estimate in $b_1$. Specifically, the coefficient $b_1$ on $T_{i1}$ captures the association flowing along the open paths from $T_{i1}$ to $D_i$. There are five paths from $T_{i1}$ to $D_i$ (listed together with their corresponding path coefficients):

\begin{enumerate}
\item $T_{i1}\leftarrow U_i\rightarrow T_{i2}\rightarrow Y_{i2}\rightarrow D_i$ (non-causal): $\delta\chi\gamma$
\item $T_{i1}\leftarrow U_i\rightarrow Y_{i1}\rightarrow D_i$ (non-causal): $-\psi\chi$
\item $T_{i1}\leftarrow U_i\rightarrow Y_{i2}\rightarrow D_i$ (non-causal): $\psi\chi$
\item $T_{i1}\rightarrow Y_{i1}\rightarrow D_i$ (causal): $\delta$
\item $T_{i1}\rightarrow Y_{i2}\rightarrow D_i$ (causal): $\theta$
\end{enumerate}

The first path is closed because the regression conditions on $T_{i2}$. The second and third paths cancel each other out exactly. The fourth path transmits the spillover effect. The fifth path transmits the negative of the targeted effect. Hence, the regression coefficient $b_1=\theta-\delta$ identifies the difference between the spillover and targeted effect.

The coefficient $b_2$ on $T_{i2}$ captures the association flowing along the open paths from $T_{i2}$ and $D_i$. There are four paths from $T_{i2}$ to $D_i$: 

\begin{enumerate}
\item $T_{i2}\gets U_i\rightarrow T_{i1}\rightarrow Y_{i1}\rightarrow D_i$ (non-causal): $-\delta\chi\gamma$
\item $T_{i2}\gets U_i\rightarrow Y_{i1}\rightarrow D_i$ (non-causal): $-\psi\gamma$
\item $T_{i2}\gets U_i\rightarrow Y_{i2}\rightarrow D_i$ (non-causal): $\psi\gamma$
\item $T_{i2}\rightarrow Y_{i2}\rightarrow D_i$ (causal): $\delta$
\end{enumerate}

The first path is closed because the regression conditions on $T_{i1}$; the second and third paths cancel each other out; and the fourth path captures the targeted effect. Thus, $b_2=\delta$ identifies the targeted effect, and $SC=b_1+b_2=\ \theta$ identifies the causal spillover effect. 

Many statistical software have functions for summing regression coefficients and obtaining standard errors. Examples include Stata’s lincom command,\cite{stataman} R’s contrast package,\cite{kuhn2016} and SAS’s SCORE procedure.\cite{sas}

The analysis is only slightly complicated in the presence of exposure-to-exposure spillover ($T_{ij}\rightarrow T_{ij^\prime}$) – for example, when one child’s serious illness increases their sibling’s risk of illness. When $T_{i2}\rightarrow T_{i1}$ (\textbf{Figure 1B}), the analysis does not change. However, if $T_{i1}\rightarrow T_{i2}$ (\textbf{Figure 1C}), then the interpretation of $SC=\theta$ changes from representing the entire spillover effect of $T_{i1}$ on $Y_{i2}$ to capturing only the direct spillover effect, since the indirect component of the spillover effect that operates via the causal path $T_{i1}\rightarrow T_{i2}\rightarrow Y_{i2}$ is closed because the regression controls for $T_{i2}$. See the \textbf{Supplementary Material} for details.

\subsubsection*{\textit{Settings with two-sided spillover}}

Analysts may also encounter scenarios with two-sided spillover. In our example, each siblings’ health shock could affect the other’s academic performance ($T_{i1}\rightarrow Y_{i2}$ and $T_{i2}\rightarrow Y_{i1}$). Reflecting this possibility, \textbf{Figure 2A} modifies the baseline model of \textbf{Figure 1A} to allow spillover $T_{i2}\rightarrow Y_{i1}$ with effect $\kappa$. The partial regression coefficients in the gain-score approach identify $b_1=\theta-\delta$ and $b_2=\delta-\kappa$, so that $SC=b_1+b_2=\theta-\kappa$. Consequently, with two-sided exposure-to-outcome spillover, the $SC$ does not identify the spillover effect of $T_{i1}$ on $Y_{i2}$ but instead the difference between the two exposure-to-outcome spillover effects. However, if the analyst can defend assumptions about one or more of the signs of the two spillover effects, then the $SC$ remains informative even though it no longer point-identifies $\theta$. Specifically, if $\kappa>0$, the $SC$ underestimates (i.e., gives a lower bound for) $\theta$. By contrast, if $\kappa<0$, then the $SC$ overestimates (gives an upper bound for) $\theta$. One can make additional inferences about $\theta$ depending on the value of the $SC$ and the assumed sign of $\kappa$. For example, if $SC>0$ and $\kappa>0$, then $\theta>0$. Of note, a finding that $SC=0$ is uninformative, because it is compatible with the possibility that the two spillover effects are equal, $\theta=\kappa$, and that they are both zero, $\theta=\kappa=0$. 

\afterpage{%
\begin{figure}
\centering
\resizebox{\linewidth}{!}{
\begin{tikzpicture}



\node [scale=1.3] (1) {U$_{i}$};

\node [above right =0.7cm and 1cm of 1, scale=1.3] (2) {T$_{i1}$};
\node [right =3cm of 2, scale=1.3] (3) {Y$_{i1}$};

\node [below right =0.7cm and 1cm of 1, scale=1.3] (4) {T$_{i2}$};
\node [right =3cm of 4, scale=1.3] (5) {Y$_{i2}$};

\node [right =2cm of 5, scale=1.3] (8) {D$_{i}$};

\node [above left =2.8cm and 0.2cm of 1, scale=1.3] (991) {\textbf{(2A)}};


\draw[Arrow] (1.east) -- (2.south west) node [midway, above, scale=1.3] {$\chi$};
\draw[Arrow3] (1) to [out=90, in=130] node [pos=0.6, above, scale=1.3] {$\psi$} (3);
\draw[Arrow3] (1) to [out=270, in=230] node [pos=0.6, below, scale=1.3] {$\psi$} (5);
\draw[Arrow] (1.east) -- (4.north west) node [midway, below, scale=1.3] {$\gamma$};

\draw[Arrow] (2.east) -- (3.west) node [midway, above, scale=1.3] {$\delta$};

\draw[Arrow] (2.south east) -- (5.north west)  node [pos=0.82, above, scale=1.3] {$\theta$};

\draw[Arrow] (4.east) -- (5.west) node [midway, above, scale=1.3] {$\delta$};

\draw[Arrow] (3.south east) -- (8.north west) node [midway, above, scale=1.3] {\it -1};
\draw[Arrow] (5.east) -- (8.west) node [midway, above, scale=1.3] {\it +1};

\draw[Arrow] (4.north east) -- (3.south west) node [pos=0.82, below, scale=1.3] {$\kappa$};




\node [right =10cm of 1, scale=1.3]  (21) {U$_{i}$};

\node [above left =2.8cm and 0.2cm of 21, scale=1.3] (993) {\textbf{(2B)}};

\node [above right =0.7cm and 1cm of 21, scale=1.3] (22) {T$_{i1}$};
\node [right =3cm of 22, scale=1.3] (23) {Y$_{i1}$};

\node [below right =0.7cm and 1cm of 21, scale=1.3] (24) {T$_{i2}$};
\node [right =3cm of 24, scale=1.3] (25) {Y$_{i2}$};

\node [right =2cm of 25, scale=1.3] (28) {D$_{i}$};


\draw[Arrow] (21.east) -- (22.south west) node [midway, above, scale=1.3] {$\chi$};
\draw[Arrow3] (21) to [out=90, in=130] node [pos=0.6, above, scale=1.3] {$\psi$} (23);
\draw[Arrow3] (21) to [out=270, in=230] node [pos=0.6, below, scale=1.3] {$\psi$} (25);
\draw[Arrow] (21.east) -- (24.north west) node [midway, below, scale=1.3] {$\gamma$};

\draw[Arrow] (22.east) -- (23.west) node [midway, above, scale=1.3] {$\delta$};

\draw[Arrow] (22.south east) -- (25.north west)  node [pos=0.82, above, scale=1.3] {$\theta$};

\draw[Arrow] (24.east) -- (25.west) node [midway, above, scale=1.3] {$\delta$};

\draw[Arrow] (23.south east) -- (28.north west) node [midway, above, scale=1.3] {\it -1};
\draw[Arrow] (25.east) -- (28.west) node [midway, above, scale=1.3] {\it +1};

\draw[Arrow] (24.north east) -- (23.south west) node [pos=0.82, below, scale=1.3] {$\kappa$};

\draw[Arrow] (24.north) -- (22.south) node [midway, left, scale=1.3] {$\tau$};




\node [below right =7cm and 5cm of 1, scale=1.3]  (41) {U$_{i}$};

\node [above left =2.8cm and 0.2cm of 41, scale=1.3] (995) {\textbf{(2C)}};

\node [above right =0.7cm and 1cm of 41, scale=1.3] (42) {T$_{i1}$};
\node [right =3cm of 42, scale=1.3] (43) {Y$_{i1}$};

\node [below right =0.7cm and 1cm of 41, scale=1.3] (44) {T$_{i2}$};
\node [right =3cm of 44, scale=1.3] (45) {Y$_{i2}$};

\node [right =2cm of 45, scale=1.3] (48) {D$_{i}$};


\draw[Arrow] (41.east) -- (42.south west) node [midway, above, scale=1.3] {$\chi$};
\draw[Arrow3] (41) to [out=90, in=130] node [pos=0.6, above, scale=1.3] {$\psi$} (43);
\draw[Arrow3] (41) to [out=270, in=230] node [pos=0.6, below, scale=1.3] {$\psi$} (45);
\draw[Arrow] (41.east) -- (44.north west) node [midway, below, scale=1.3] {$\gamma$};

\draw[Arrow] (42.east) -- (43.west) node [midway, above, scale=1.3] {$\delta$};

\draw[Arrow] (42.south east) -- (45.north west)  node [pos=0.82, above, scale=1.3] {$\theta$};

\draw[Arrow] (44.east) -- (45.west) node [midway, above, scale=1.3] {$\delta$};

\draw[Arrow] (43.south east) -- (48.north west) node [midway, above, scale=1.3] {\it -1};
\draw[Arrow] (45.east) -- (48.west) node [midway, above, scale=1.3] {\it +1};

\draw[Arrow] (44.north east) -- (43.south west) node [pos=0.82, below, scale=1.3] {$\kappa$};

\draw[Arrow] (42.south) -- (44.north) node [midway, left, scale=1.3] {$\phi$};

\end{tikzpicture}
}
\caption{\textbf{Figure 2.} Causal directed acyclic graphs for linear data-generating models with two-sided exposure-to-outcome sibling spillover. Subscripts $i$ and $j$ denote cluster and sibling, respectively. $T_{ij}$ is the exposure, $Y_{ij}$ is the outcome, $D_i$ is the gain-score, and $U_i$ is an unobserved family-level confounder. Greek letters denote effects. (2A) does not have exposure-to-exposure spillover, whereas (2B) and (2C) have exposure-to-exposure spillover.} 
\end{figure}
\clearpage
}

If $T_{ij}\rightarrow T_{ij'}$ in addition to two-sided spillover (\textbf{Figures 2B-C}), this does not affect the interpretations of $b_1$ and $b_2$, and $SC$ still identifies the difference between siblings’ unmediated spillover effects. However, $SC$ will not capture the mediated part of spillover effect, $T_{ij}\rightarrow T_{ij'}\rightarrow Y_{ij'}$. See the \textbf{Supplementary Material} for details.

\subsubsection*{\textit{Settings with spillovers from outcomes}}

Analysts may also encounter settings with outcome-to-outcome spillover ($Y_{ij}\rightarrow Y_{ij'}$) or outcome-to-exposure spillover ($Y_{ij}\rightarrow T_{ij'}$). In our setting, it is reasonable to assume that siblings’ academic outcomes may be causally related by outcome-to-outcome spillover. In contrast, an academic outcome causing a health shock is implausible, but exposure-to-outcome spillovers may be relevant elsewhere.

If outcomes cause future exposures or outcomes (\textbf{Figure 3}), then our estimator does not identify spillovers or simple functions of spillovers. See the \textbf{Supplementary Material} for details.

\afterpage{%
\begin{figure}
\centering
\resizebox{\linewidth}{!}{
\begin{tikzpicture}



\node [scale=1.3] (1) {U$_{i}$};

\node [above right =0.7cm and 1cm of 1, scale=1.3] (2) {T$_{i1}$};
\node [right =3cm of 2, scale=1.3] (3) {Y$_{i1}$};

\node [below right =0.7cm and 1cm of 1, scale=1.3] (4) {T$_{i2}$};
\node [right =3cm of 4, scale=1.3] (5) {Y$_{i2}$};

\node [right =2cm of 5, scale=1.3] (8) {D$_{i}$};

\node [above left =2.8cm and 0.2cm of 1, scale=1.3] (991) {\textbf{(3A)}};


\draw[Arrow] (1.east) -- (2.south west) node [midway, above, scale=1.3] {$\chi$};
\draw[Arrow3] (1) to [out=90, in=130] node [pos=0.6, above, scale=1.3] {$\psi$} (3);
\draw[Arrow3] (1) to [out=270, in=230] node [pos=0.6, below, scale=1.3] {$\psi$} (5);
\draw[Arrow] (1.east) -- (4.north west) node [midway, below, scale=1.3] {$\gamma$};

\draw[Arrow] (2.east) -- (3.west) node [midway, above, scale=1.3] {$\delta$};

\draw[Arrow] (2.south east) -- (5.north west)  node [pos=0.82, above, scale=1.3] {$\theta$};

\draw[Arrow] (4.east) -- (5.west) node [midway, above, scale=1.3] {$\delta$};

\draw[Arrow] (3.south east) -- (8.north west) node [midway, above, scale=1.3] {\it -1};
\draw[Arrow] (5.east) -- (8.west) node [midway, above, scale=1.3] {\it +1};

\draw[Arrow] (3.south west) -- (4.north east) node [pos=0.82, above, scale=1.3] {$\omega$};




\node [right =10cm of 1, scale=1.3]  (21) {U$_{i}$};

\node [above left =2.8cm and 0.2cm of 21, scale=1.3] (993) {\textbf{(3B)}};

\node [above right =0.7cm and 1cm of 21, scale=1.3] (22) {T$_{i1}$};
\node [right =3cm of 22, scale=1.3] (23) {Y$_{i1}$};

\node [below right =0.7cm and 1cm of 21, scale=1.3] (24) {T$_{i2}$};
\node [right =3cm of 24, scale=1.3] (25) {Y$_{i2}$};

\node [right =2cm of 25, scale=1.3] (28) {D$_{i}$};


\draw[Arrow] (21.east) -- (22.south west) node [midway, above, scale=1.3] {$\chi$};
\draw[Arrow3] (21) to [out=90, in=130] node [pos=0.6, above, scale=1.3] {$\psi$} (23);
\draw[Arrow3] (21) to [out=270, in=230] node [pos=0.6, below, scale=1.3] {$\psi$} (25);
\draw[Arrow] (21.east) -- (24.north west) node [midway, below, scale=1.3] {$\gamma$};

\draw[Arrow] (22.east) -- (23.west) node [midway, above, scale=1.3] {$\delta$};

\draw[Arrow] (22.south east) -- (25.north west)  node [midway, above, scale=1.3] {$\theta$};

\draw[Arrow] (24.east) -- (25.west) node [midway, above, scale=1.3] {$\delta$};

\draw[Arrow] (23.south east) -- (28.north west) node [midway, above, scale=1.3] {\it -1};
\draw[Arrow] (25.east) -- (28.west) node [midway, above, scale=1.3] {\it +1};

\draw[Arrow] (23.south) -- (25.north) node [midway, left, scale=1.3] {$\eta$};




\node [below right =7cm and 5cm of 1, scale=1.3]  (41) {U$_{i}$};

\node [above left =2.8cm and 0.2cm of 41, scale=1.3] (995) {\textbf{(3C)}};

\node [above right =0.7cm and 1cm of 41, scale=1.3] (42) {T$_{i1}$};
\node [right =3cm of 42, scale=1.3] (43) {Y$_{i1}$};

\node [below right =0.7cm and 1cm of 41, scale=1.3] (44) {T$_{i2}$};
\node [right =3cm of 44, scale=1.3] (45) {Y$_{i2}$};

\node [right =2cm of 45, scale=1.3] (48) {D$_{i}$};


\draw[Arrow] (41.east) -- (42.south west) node [midway, above, scale=1.3] {$\chi$};
\draw[Arrow3] (41) to [out=90, in=130] node [pos=0.6, above, scale=1.3] {$\psi$} (43);
\draw[Arrow3] (41) to [out=270, in=230] node [pos=0.6, below, scale=1.3] {$\psi$} (45);
\draw[Arrow] (41.east) -- (44.north west) node [midway, below, scale=1.3] {$\gamma$};

\draw[Arrow] (42.east) -- (43.west) node [midway, above, scale=1.3] {$\delta$};

\draw[Arrow] (42.south east) -- (45.north west)  node [midway, above, scale=1.3] {$\theta$};

\draw[Arrow] (44.east) -- (45.west) node [midway, above, scale=1.3] {$\delta$};

\draw[Arrow] (43.south east) -- (48.north west) node [midway, above, scale=1.3] {\it -1};
\draw[Arrow] (45.east) -- (48.west) node [midway, above, scale=1.3] {\it +1};

\draw[Arrow] (45.north) -- (43.south) node [midway, left, scale=1.3] {$\lambda$};

\end{tikzpicture}
}
\caption{\textbf{Figure 3.} Causal directed acyclic graphs for linear data-generating models with one-sided exposure-to-outcome sibling spillover and spillover from outcomes. Subscripts $i$ and $j$ denote cluster and sibling, respectively. $T_{ij}$ is the exposure, $Y_{ij}$ is the outcome, $D_i$ is the gain-score, and $U_i$ is an unobserved family-level confounder. Greek letters denote effects. (3A) has outcome-to-exposure spillover, and (3B) and (3C) have outcome-to-outcome spillover.} 
\end{figure}
\clearpage
}


\section{Simulation}

We conducted nine Monte Carlo simulations,\cite{adkins2012} one simulation for each of the nine models in \textbf{Figures 1-3}, to demonstrate when the method identifies exposure-to-outcome spillover effects. Our simulation model follows:

\[U_i,\upsilon_{i1},\upsilon_{i2}\sim N(0,1)\]
\[T_{i1}=
	\begin{cases}
	0 \ \text{if}\ \tau T_{i2}+\chi U_i\leq 0.5 \\
	1 \ \text{if}\ \tau T_{i2}+\chi U_i > 0.5
	\end{cases}\]
\[T_{i2}=
	\begin{cases}
	0 \ \text{if}\ \phi T_{i1}+\omega Y_{i1}+\gamma U_i\leq 0.2 \\
	1 \ \text{if}\ \phi T_{i1}+\omega Y_{i1}+\gamma U_i > 0.2
	\end{cases}\]
\[Y_{i1}=\delta T_{i1}+\kappa T_{i2}+\lambda Y_{i2}+\psi U_i+\upsilon_{i1}\]
\[Y_{i2}=\delta T_{i2}+\theta T_{i1}+\eta Y_{i1}+\psi U_i+\upsilon_{i2}\]
\[D_i=Y_{i2}-Y_{i1}\]

We simulated each model with 1000 runs of 5000 observations each, where each observation represented a sibling pair. We set the following parameters at fixed values: $\theta=0.5$, $\delta=1$, $\psi=1$, $\chi=2$, and $\gamma=3$. Parameters distinguishing the models---$\kappa$, $\tau$, $\phi$, $\omega$, $\eta$, and $\lambda$---were set to zero unless otherwise specified. To avoid simultaneity, at least one parameter in each pair ($\tau$, $\phi$), ($\eta$, $\lambda$), and ($\kappa$, $\omega$) was always set to zero. In each sample, we regressed the gain-score on siblings’ exposures and computed the spillover coefficient according to equations (1) and (2). We conducted simulations in Stata Statistical Software: Release 16.\cite{statasoft} Simulation code is in the \textbf{Supplementary Material}. 

\textbf{Figure 4} displays the simulation results. The first three rows confirm that the spillover coefficient is unbiased in the three settings with one-sided exposure-to-outcome spillover of \textbf{Figure 1}, as the average of estimated spillover coefficient equals the known spillover effect, ${\widehat{SC}}_{Figure1}=0.5$ (empirical 95\% CI: 0.42, 0.58). The subsequent three rows demonstrate that the spillover coefficient in the three models of \textbf{Figure 2} with two-sided exposure-to-outcome spillover identifies the difference between the two spillovers, ${\widehat{SC}}_{Figure2}=0.5-0.3=0.2$ (empirical 95\% CI: 0.12, 0.28). Since $\kappa>0$, ${\widehat{SC}}_{Figure2}$ underestimates the spillover effect, $\theta$. The final three simulations show that ${\widehat{SC}}_{Figure3}$ is biased in all models of \textbf{Figure 3} with spillovers from outcomes. Size and direction of the biases are fairly complicated functions of the coefficients in the data-generating model and can be large. The estimated ordinary least squares standard errors closely resemble the empirical standard errors for each model, indicating that the built-in standard errors in Stata’s lincom command are accurate.\cite{stataman} 

\afterpage{%
\begin{figure}
\centering
\includegraphics{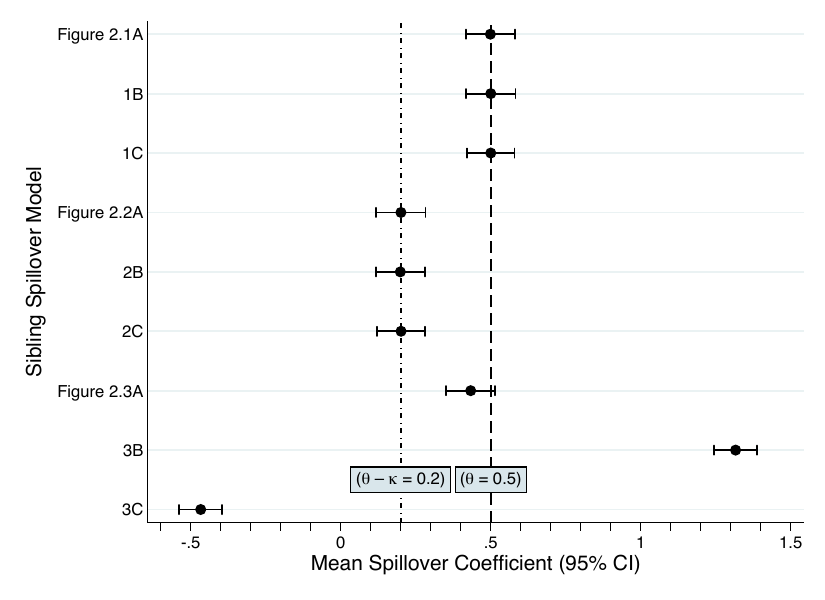}
\caption{\textbf{Figure 4.} Results (average spillover coefficients and empirical 95\% confidence intervals [CI]) from simulations of the nine sibling spillover models in \textbf{Figures 1-3}. Each simulation consisted of 1000 runs of 5000 observations, where each observation represented a sibling pair. Subscripts $i$ and $j$ indicate cluster and sibling, respectively. $T_{ij}$ is the exposure and $Y_{ij}$ is the outcome. The target quantity is the spillover effect ($T_{i1} \rightarrow Y_{i2}$), set to $\theta=0.5$ in all models. Other spillover effects include $\kappa$ ($T_{i2} \rightarrow Y_{i1}$), $\tau$ ($T_{i2} \rightarrow T_{i1}$), $\phi$ ($T_{i1} \rightarrow T_{i2}$), $\omega$ ($Y_{i1} \rightarrow T_{i2}$), $\eta$ ($Y_{i1} \rightarrow Y_{i2}$), and $\lambda$ ($Y_{i2} \rightarrow Y_{i1}$). Except for $\theta$, all spillover parameters were set to zero except in the following cases: $\kappa=0.3$ in 2A-C; $\tau=0.3$ in 1B and 2B; $\phi=0.3$ in 1C and 2C; $\omega=0.3$ in 3A; $\eta=0.3$ in 3B; and $\lambda=0.3$ in 3C. The spillover coefficient identifies (is unbiased) for $\theta$ in all models of one-sided spillover (\textbf{Figure 1}); identifies the difference between the two spillover effects with two-sided spillover (\textbf{Figure 2}); and is biased in the presence of spillovers originating from outcomes (\textbf{Figure 3}).} 
\end{figure}
\clearpage
}


\section{Empirical Application}

We applied the method to estimating the spillover effect of a child’s preterm birth (gestational age $<$ 37 weeks) on their older sibling’s literacy skills. This analysis builds upon evidence that short gestational age and health shocks within the family may impede a child's own early literacy skills.\cite{black2020, mathiasen2010, mallinson2019} If a child is born preterm, parents may reallocate investments (time, financial, or otherwise) from older siblings to support the younger sibling's health, thereby inhibiting the older siblings' development, including early literacy.

For this application, we analyzed Big Data for Little Kids (BD4LK), a longitudinal cohort of birth records for all live in-state resident deliveries in Wisconsin during 2007-2016 (N $>$ 660,000 deliveries) that links to multiple administrative data sources, including Medicaid data (2007-2016) and children’s Phonological Awareness Literacy Screening-Kindergarten (PALS-K) test scores from Wisconsin public schools (2012-2016 school years). BD4LK’s linking process is described elsewhere.\cite{mallinson2019, larson2019} PALS-K evaluates readiness for kindergarten-level literacy instruction on six domains (rhyme awareness; beginning sound awareness; alphabet knowledge; letter sounds; spelling; word concept).\cite{ford2014, pktech} In Wisconsin, children must be five years-old at kindergarten enrollment to qualify for PALS-K testing.\cite{widpi} Our analysis includes 20,010 sibling pairs (40,020 children) that were sequentially-born from different deliveries to the same biological mother and had non-missing English-language PALS-K test scores and covariates. The \textbf{Supplementary Material} contains the full sampling description. 

We estimate the following gain-score regression model, 

\[D_i=b_1{PTB}_{i1}+b_2{PTB}_{i2}+\beta_3\textbf{\textit{C}}_{i2}+v_i\]

\noindent where $D_i=PALSK_{i2}-PALSK_{i1}$. Subscripts $i=1,\ldots N$ and $j=1,2$ indicate cluster and sibling, respectively, where $j=1$ is the younger sibling. $PTB_{ij}$ is a binary preterm birth indicator (1 if preterm; 0 otherwise), $PALSK_{ij}$ is the continuous PALS-K score (0-102 points), and $\textbf{\textit{C}}_{i2}$ is a vector of covariates measured at the older sibling’s delivery, which may be empty. Covariates include maternal age (years), maternal education (no high school diploma; high school diploma/equivalent; 1-3 years college; 4+ years college), and Medicaid delivery payment. $D_i$ is the gain-score estimator. 

We ran the model twice, once with and once without covariates. In each model, we summed the regression coefficients on both siblings’ preterm birth indicators to compute the $SC=b_1+b_2$. Assuming the one-sided spillover model of \textbf{Figure 1A}, $SC$ from the regression without covariates identifies the effect of a younger sibling’s preterm birth on the older sibling’s PALS-K score. Additionally, $b_2$ identifies the effect of each sibling’s preterm birth on their own PALS-K score. We performed all analyses in Stata Statistical Software: Release 16.\cite{statasoft} The University of Wisconsin-Madison minimal risk institutional review board approved our project. 

\textbf{Supplemental Tables 1 and 2} summarize baseline characteristics of our sample (\textbf{Supplementary Material}). Preterm birth incidence was slightly greater among older siblings relative to younger siblings (6.78\% vs. 6.65\%). On average, older siblings received slightly lower PALS-K scores (mean 63.58 points; SD 24.12 points) relative to younger siblings (mean 64.22 points; SD 23.83 points). Approximately 10\% of sibling clusters had discordant preterm birth exposure. In the regression without covariate adjustment, the older sibling's preterm birth coefficient was ${\hat{b}}_2$ = -2.49 points (95\% CI -3.83, -1.15 points), the younger sibling's preterm birth coefficient was ${\hat{b}}_1$ = 0.38 points (95\% CI: -0.97, 1.73 points), and the resulting $\widehat{SC}$ was -2.11 points (95\% CI: -3.82, -0.40 points) (\textbf{Table 1}). This indicates that a younger sibling’s preterm birth modestly harmed their older sibling’s PALS-K performance. \textbf{Figure 5} displays these results graphically relative to the assumed data-generating model. However, covariate adjustment attenuated the $\widehat{SC}$ to -1.49 points (95\% CI -3.21, 0.22 points).

\afterpage{%
\begin{table}[ht]
\setlength\extrarowheight{5pt}
\centering
\captionsetup{justification=justified, singlelinecheck=false}
\caption{\textbf{Table 1.} Ordinary least squares regression of the difference in siblings’ PALS-K scores$^a$ (points) on their preterm birth statuses (N = 20,010 sibling pairs) }
\begin{adjustbox}{max height=\textheight}{%
\begin{tabular} { l r r }
\hline
{} & \textbf{Unadjusted Regression} & \textbf{Adjusted Regression}$^b$ \\
{} & \textbf{Coefficient (95\% CI)} & \textbf{Coefficient (95\% CI)} \\
\hline
Preterm birth (gestational age $<$ 37 weeks) & {} & {} \\
\textit{Older sibling} & -2.49 (-3.83, -1.15) & -2.28 (-3.62, -0.94) \\
\textit{Younger sibling} & 0.38 (-0.97, 1.73) & 0.79 (-0.57, 2.14) \\
\hline
Spillover coefficient$^c$ & -2.11 (-3.82, -0.40) & -1.49 (-3.21, 0.22) \\
\hline
\end{tabular}
}
\end{adjustbox}
\setlength\extrarowheight{0pt}
\caption{$^{a}$The difference in PALS-K scores equals the older sibling’s PALS-K Score minus the younger sibling’s PALS-K score.}
\caption{$^{b}$Covariates include  maternal age at delivery (years), maternal education at delivery (no high school diploma; high school diploma/equivalent; 1-3 years college; 4+ years college) and Medicaid delivery payment (no; yes), all of which were measured at the time of the older sibling’s delivery.} 
\caption{$^{c}$The spillover coefficient is the sum of the partial regression coefficients for the older sibling’s preterm birth indicator and the younger sibling’s preterm birth indicator. Assuming one-sided spillover as in \textbf{Figure 1A}, this identifies the effect of a younger sibling’s preterm birth on the older sibling’s PALS-K score.} 
\caption{Abbreviations: ``CI” confidence interval; ``PALS-K” Phonological Awareness Literacy Screening-Kindergarten.}
\end{table}
\clearpage
}

\afterpage{%
\begin{figure}
\centering
\resizebox{\linewidth}{!}{
\begin{tikzpicture}


\node (1) {U$_{i}$};

\node [above right =0.7cm and 1cm of 1] (2) {PTB$_{i1}$};
\node [right =4.5cm of 2] (3) {PALSK$_{i1}$};

\node [below right =0.7cm and 1cm of 1] (4) {PTB$_{i2}$};
\node [right =4.5cm of 4] (5) {PALSK$_{i2}$};

\node [right =2cm of 5] (8) {D$_{i}$};



\draw[Arrow] (1.east) -- (2.south west) node [midway, above] {$\chi$};
\draw[Arrow3] (1) to [out=90, in=130] node [pos=0.6, above] {$\psi$} (3);
\draw[Arrow3] (1) to [out=270, in=230] node [pos=0.6, below] {$\psi$} (5);
\draw[Arrow] (1.east) -- (4.north west) node [midway, below] {$\gamma$};

\draw[Arrow] (2.east) -- (3.west) node [midway, above] {$\hat{\delta} = -2.49$};

\draw[Arrow] (2.south east) -- (5.north west) node [midway, above, rotate=338] {$\hat{\theta} = -2.11$};

\draw[Arrow] (4.east) -- (5.west) node [midway, above] {$\hat{\delta} = -2.49$};

\draw[Arrow] (3.south east) -- (8.north west) node [midway, above] {\it -1};
\draw[Arrow] (5.east) -- (8.west) node [midway, above] {\it +1};

\end{tikzpicture}
}
\caption{\textbf{Figure 5.} A directed acyclic graph of the relationship between siblings’ preterm birth (gestational age $<$ 37 weeks) and their score on the Phonological Awareness Literacy Assessment-Kindergarten test with overlaid estimates. Subscripts $i$ and $j$ denote cluster and sibling, respectively, where $j=1$ is the younger sibling and $j=2$ is the older sibling. $PTB_{ij}$ is a preterm birth indicator, $PALSK_{ij}$ is the test score, $D_i$ is a gain-score, and $U_i$ is an unobserved confounder. Greek letters denote effects, and the values of $\theta$ and $\delta$ are estimated using gain-score regression.} 
\end{figure}
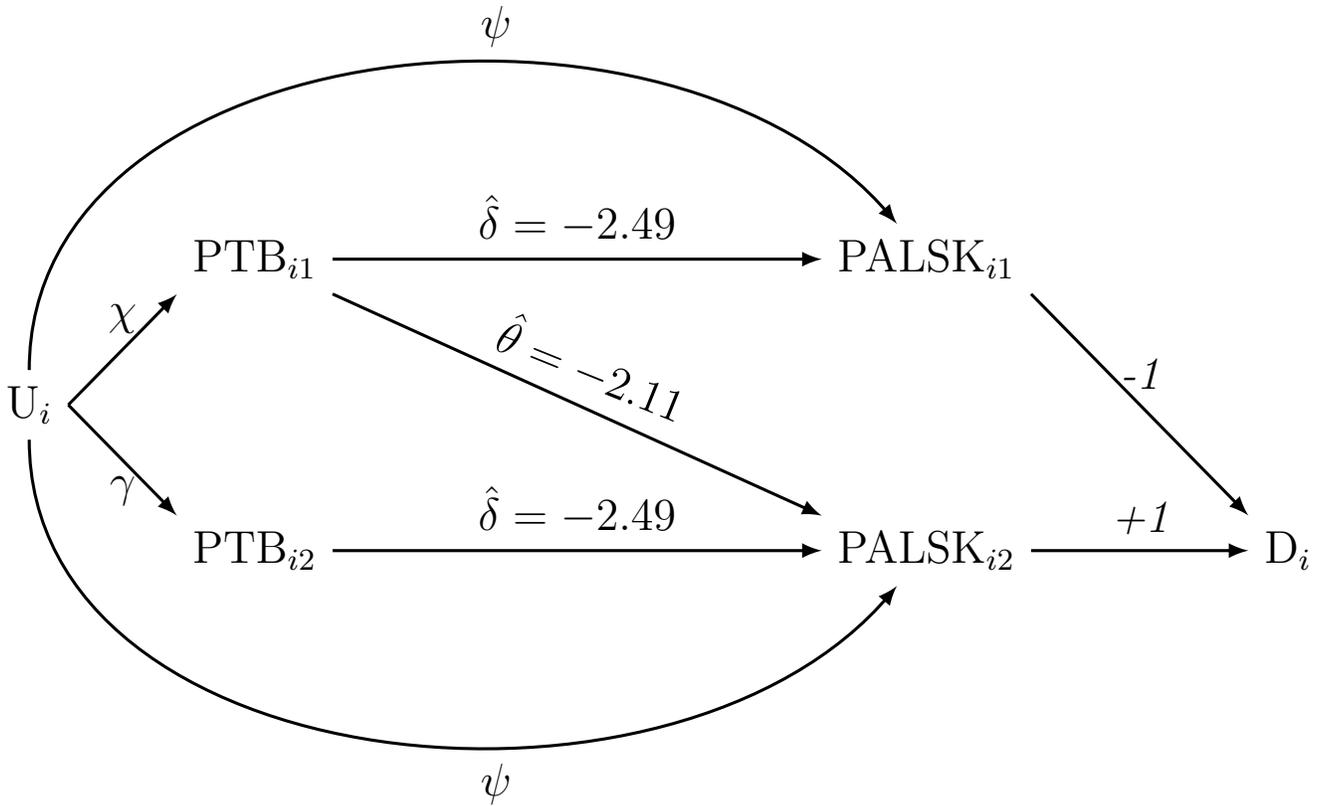
\clearpage
}


\section{Discussion}

We described a simple approach to identifying spillovers with gain-scores in sibling pairs. This method can point-identify spillovers if only one sibling’s exposure affects the other’s outcome, and it can identify the difference in siblings’ spillovers in the presence of two-sided spillover. The method leverages the primary benefit of FE estimation: controlling for family-level, sibling-invariant, unobserved confounding. Whereas preceding epidemiologic research on spillover identification primarily considered infectious diseases, our work contributes to the growing literature on spillovers within families.

We acknowledge some limitations. First, we restricted our attention to linear settings. This method does not necessarily apply to contexts with nonlinear relationships, such as those with binary outcomes (see Sjölander et al. (2016)\cite{sjolander2016} for binary outcomes in our \textbf{Figure 1A}). Second, we did not consider clusters of three or more siblings. Spillovers that originate from larger sibling clusters may pose unique challenges that are unaddressed here – for example, whether one can identify the effect of a middle child's exposure on the youngest sibling's outcome if an eldest sibling's exposure affects all siblings' outcomes. Lastly, we did not test the method in the presence of shared mediator or collider variables. Sjölander and Zetterqvist (2017) interrogated sibling comparison models with shared mediators and colliders, finding that such factors may induce bias.\cite{sjolander2017} 

Nonetheless, our paper lays groundwork for subsequent research. Specific avenues that advance this method include testing in nonlinear settings or settings with shared mediator variables, as well as expanding models to allow three or more siblings. 

\newpage

\newpage


\appendix

\begin{titlepage}
  \centering
  \vspace*{\fill}
  \vskip 60pt
  \LARGE Estimating Sibling Spillover Effects with Unobserved Confounding Using Gain-Scores \\
 \textit{Supplementary Material} \par
  \vskip 3em
  \large 
  \vspace*{\fill}
\end{titlepage}

\small{}

\renewcommand{\thepage}{S\arabic{page}}


\section{Derivations for Models with One-Sided Spillover}

\subsection*{\textit{Base model with one-sided spillover (Figure 1A)}}

We first consider models with one-sided exposure-to-outcome spillover ($T_{i1} \rightarrow Y_{i2}$, but $T_{i2}$ does not directly affect $Y_{i1}$). Under the data generating process in \textbf{Figure 1A}, we point identify the spillover effect $\theta$. For reference, $\beta$ denotes a regression coefficient, $\rho$ denotes a correlation coefficient, $\sigma$ denotes standard deviation, and $\sigma^2$ denotes variance. We compute the following partial regression coefficients: 

\begin{flalign*}
\beta_{D_{i}T_{i1} \cdot T_{i2}} &= \frac{\rho_{D_{i}T_{i1}}-\rho_{D_{i}T_{i2}}*\rho_{T_{i1}T_{i2}}}{1-{\rho_{T_{i1}T_{i2}}}^2} 
*
\frac{\sigma_{D_{i}}}{\sigma_{T_{i1}}}\\ 
\end{flalign*}
\begin{flalign*}
\beta_{D_{i}T_{i2} \cdot T_{i1}} &= \frac{\rho_{D_{i}T_{i2}}-\rho_{D_{i}T_{i1}}*\rho_{T_{i1}T_{i2}}}{1-{\rho_{T_{i1}T_{i2}}}^2} 
*
\frac{\sigma_{D_{i}}}{\sigma_{T_{i2}}}\\ 
\end{flalign*}

\noindent The correlation coefficient between two variables is equal to the sum of the products of path coefficients and of the "root variable's" variation divided by the product the variables' standard deviations. A "root variable" is the variable on the path with no incoming arrows. See Pearl (2013) for details.\cite{pearl2013} We compute $\rho_{D_{i}T_{i1}}$, $\rho_{D_{i}T_{i2}}$, and $\rho_{T_{i1}T_{i2}}$:

\begin{flalign*}
\rho_{D_{i}T_{i1}} &= \frac{(-1){{\sigma_{U_{i}}}^2}\psi\chi+(-1){{\sigma_{T_{i1}}}^2}\delta+{{\sigma_{T_{i1}}}^2}\theta+{{\sigma_{U_{i}}}^2}\delta\chi\gamma+{{\sigma_{U_{i}}}^2}\psi\chi}{\sigma_{D_{i}}\sigma_{T_{i1}}} \\ 
	&= \frac{-{{\sigma_{T_{i1}}}^2}\delta+{{\sigma_{T_{i1}}}^2}\theta+{{\sigma_{U_{i}}}^2}\delta\chi\gamma}{\sigma_{D_{i}}\sigma_{T_{i1}}}
\end{flalign*}
\begin{flalign*}
\rho_{D_{i}T_{i2}} &= \frac{(-1){{\sigma_{U_{i}}}^2}\psi\gamma+(-1){{\sigma_{U_{i}}}^2}\delta\chi\gamma+{{\sigma_{U_{i}}}^2}\theta\chi\gamma+{{\sigma_{T_{i2}}}^2}\delta+{{\sigma_{U_{i}}}^2}\psi\gamma}{\sigma_{D_{i}}\sigma_{T_{i2}}} \\ 
	&= \frac{-{{\sigma_{U_{i}}}^2}\delta\chi\gamma+{{\sigma_{U_{i}}}^2}\theta\chi\gamma+{{\sigma_{T_{i2}}}^2}\delta}{\sigma_{D_{i}}\sigma_{T_{i2}}}
\end{flalign*}
\begin{flalign*}
\rho_{T_{i1}T_{i2}} &= \frac{{{\sigma_{U_{i}}}^2}\chi\gamma}{\sigma_{T_{i1}}\sigma_{T_{i2}}} 
\end{flalign*}

\noindent We plug the correlation coefficients into the formulas for our partial regression coefficients, and we derive the following: 

\begin{flalign*}
\beta_{D_{i}T_{i1} \cdot T_{i2}} &= 
\frac{
{\frac{-{{\sigma_{T_{i1}}}^2}\delta+{{\sigma_{T_{i1}}}^2}\theta+{{\sigma_{U_{i}}}^2}\delta\chi\gamma}{\sigma_{D_{i}}\sigma_{T_{i1}}}}
-
{\frac{-{{\sigma_{U_{i}}}^2}\delta\chi\gamma+{{\sigma_{U_{i}}}^2}\theta\chi\gamma+{{\sigma_{T_{i2}}}^2}\delta}{\sigma_{D_{i}}\sigma_{T_{i2}}}}
*
{\frac{{{\sigma_{U_{i}}}^2}\chi\gamma}{\sigma_{T_{i1}}\sigma_{T_{i2}}} }
}
{
1-({\frac{{{\sigma_{U_{i}}}^2}\chi\gamma}{\sigma_{T_{i1}}\sigma_{T_{i2}}}})^2
} 
*
\frac{\sigma_{D_{i}}}{\sigma_{T_{i1}}}\\ 
    &= \theta-\delta
\end{flalign*}

\begin{flalign*}
\beta_{D_{i}T_{i2} \cdot T_{i1}} &= \frac{
{\frac{-{{\sigma_{U_{i}}}^2}\delta\chi\gamma+{{\sigma_{U_{i}}}^2}\theta\chi\gamma+{{\sigma_{T_{i2}}}^2}\delta}{\sigma_{D_{i}}\sigma_{T_{i2}}}}
-
{\frac{-{{\sigma_{T_{i1}}}^2}\delta+{{\sigma_{T_{i1}}}^2}\theta+{{\sigma_{U_{i}}}^2}\delta\chi\gamma}{\sigma_{D_{i}}\sigma_{T_{i1}}}}
*
{\frac{{{\sigma_{U_{i}}}^2}\chi\gamma}{\sigma_{T_{i1}}\sigma_{T_{i2}}} }
}
{
1-({\frac{{{\sigma_{U_{i}}}^2}\chi\gamma}{\sigma_{T_{i1}}\sigma_{T_{i2}}}})^2
} 
*
\frac{\sigma_{D_{i}}}{\sigma_{T_{i2}}}\\ 
	&= \delta
\end{flalign*}

\noindent We can then compute the spillover coefficient ($SC$) to point identify $\theta$: 
\begin{flalign*}
SC &= \beta_{D_{i}T_{i1} \cdot T_{i2}}+\beta_{D_{i}T_{i2} \cdot T_{i1}} \\
	&= \theta-\delta+\delta \\
	&= \theta \\
\end{flalign*}

\subsection*{\textit{One-sided spillover and exposure-to-exposure spillover (Figures 1B-1C)}}

Under the data generating process in \textbf{Figure 1B}, sibling 2's exposure affects sibling 1's exposure ($T_{i2} \rightarrow T_{i1}$). Regardless, we still point identify the spillover effect $\theta$. We compute $\rho_{D_{i}T_{i1}}$, $\rho_{D_{i}T_{i2}}$, and $\rho_{T_{i1}T_{i2}}$:

\begin{flalign*}
\rho_{D_{i}T_{i1}} &= \frac{\splitfrac{(-1){{\sigma_{U_{i}}}^2}\psi\chi+(-1){{\sigma_{U_{i}}}^2}\psi\gamma\tau+(-1){{\sigma_{T_{i1}}}^2}\delta}{+{{\sigma_{T_{i1}}}^2}\theta+{{\sigma_{T_{i2}}}^2}\delta\tau+{{\sigma_{U_{i}}}^2}\delta\chi\gamma+{{\sigma_{U_{i}}}^2}\psi\gamma\tau+{{\sigma_{U_{i}}}^2}\psi\chi}}{{\sigma_{D_{i}}\sigma_{T_{i1}}}} \\ 
	&= \frac{-{{\sigma_{T_{i1}}}^2}\delta+{{\sigma_{T_{i1}}}^2}\theta+{{\sigma_{T_{i2}}}^2}\delta\tau+{{\sigma_{U_{i}}}^2}\delta\chi\gamma}{{\sigma_{D_{i}}\sigma_{T_{i1}}}}
\end{flalign*}
\begin{flalign*}
\rho_{D_{i}T_{i2}} &= \frac{(-1){{\sigma_{U_{i}}}^2}\psi\gamma+(-1){{\sigma_{U_{i}}}^2}\delta\chi\gamma+(-1){{\sigma_{T_{i2}}}^2}\delta\tau+{{\sigma_{U_{i}}}^2}\theta\chi\gamma+{{\sigma_{T_{i2}}}^2}\theta\tau+{{\sigma_{T_{i2}}}^2}\delta+{{\sigma_{U_{i}}}^2}\psi\gamma}{{\sigma_{D_{i}}\sigma_{T_{i2}}}} \\ 
	&= \frac{-{{\sigma_{U_{i}}}^2}\delta\chi\gamma-{{\sigma_{T_{i2}}}^2}\delta\tau+{{\sigma_{U_{i}}}^2}\theta\chi\gamma+{{\sigma_{T_{i2}}}^2}\theta\tau+{{\sigma_{T_{i2}}}^2}\delta}{{\sigma_{D_{i}}\sigma_{T_{i2}}}}
\end{flalign*}
\begin{flalign*}
\rho_{T_{i1}T_{i2}} &= \frac{{{\sigma_{U_{i}}}^2}\chi\gamma+{{\sigma_{T_{i2}}}^2}\tau}{{\sigma_{T_{i1}}\sigma_{T_{i2}}}}
\end{flalign*}

\noindent We compute $\beta_{D_{i}T_{i1} \cdot T_{i2}}$ and $\beta_{D_{i}T_{i2} \cdot T_{i1}}$:

\begin{flalign*}
\beta_{D_{i}T_{i1} \cdot T_{i2}} &= 
{\frac{\splitfrac{
\frac{-{{\sigma_{T_{i1}}}^2}\delta+{{\sigma_{T_{i1}}}^2}\theta+{{\sigma_{T_{i2}}}^2}\delta\tau+{{\sigma_{U_{i}}}^2}\delta\chi\gamma}{{\sigma_{D_{i}}\sigma_{T_{i1}}}}}
{-
\frac{-{{\sigma_{U_{i}}}^2}\delta\chi\gamma-{{\sigma_{T_{i2}}}^2}\delta\tau+{{\sigma_{U_{i}}}^2}\theta\chi\gamma+{{\sigma_{T_{i2}}}^2}\theta\tau+{{\sigma_{T_{i2}}}^2}\delta}{{\sigma_{D_{i}}\sigma_{T_{i2}}}}
*
\frac{{{\sigma_{U_{i}}}^2}\chi\gamma+{{\sigma_{T_{i2}}}^2}\tau}{{\sigma_{T_{i1}}\sigma_{T_{i2}}}}
}}{1-(\frac{{{\sigma_{U_{i}}}^2}\chi\gamma+{{\sigma_{T_{i2}}}^2}\tau}{{\sigma_{T_{i1}}\sigma_{T_{i2}}}})^2}}
*
\frac{\sigma_{D_{i}}}{\sigma_{T_{i1}}}\\ 
    &= \theta-\delta
\end{flalign*}

\begin{flalign*}
\beta_{D_{i}T_{i2} \cdot T_{i1}} &= \frac{\splitfrac{
\frac{-{{\sigma_{U_{i}}}^2}\delta\chi\gamma-{{\sigma_{T_{i2}}}^2}\delta\tau+{{\sigma_{U_{i}}}^2}\theta\chi\gamma+{{\sigma_{T_{i2}}}^2}\theta\tau+{{\sigma_{T_{i2}}}^2}\delta}{{\sigma_{D_{i}}\sigma_{T_{i2}}}}}
{-
\frac{-{{\sigma_{T_{i1}}}^2}\delta+{{\sigma_{T_{i1}}}^2}\theta+{{\sigma_{T_{i2}}}^2}\delta\tau+{{\sigma_{U_{i}}}^2}\delta\chi\gamma}{{\sigma_{D_{i}}\sigma_{T_{i1}}}}
*
\frac{{{\sigma_{U_{i}}}^2}\chi\gamma+{{\sigma_{T_{i2}}}^2}\tau}{{\sigma_{T_{i1}}\sigma_{T_{i2}}}}
}}
{
1-(\frac{{{\sigma_{U_{i}}}^2}\chi\gamma+{{\sigma_{T_{i2}}}^2}\tau}{{\sigma_{T_{i1}}\sigma_{T_{i2}}}})^2
} 
*
\frac{\sigma_{D_{i}}}{\sigma_{T_{i2}}}\\ 
	&= \delta
\end{flalign*}

\noindent Finally, we compute $\textit{SC}$:
\begin{flalign*}
SC &= \beta_{D_{i}T_{i1} \cdot T_{i2}}+\beta_{D_{i}T_{i2} \cdot T_{i1}} \\
	&= \theta-\delta+\delta \\
	&= \theta \\
\end{flalign*}

The data generating process in \textbf{Figure 1C} also has exposure-to-exposure spillover. In this setting, sibling 1's exposure affects sibling 2's exposure ($T_{i1} \rightarrow T_{i2}$). Still, we point identify the spillover effect $\theta$. We compute $\rho_{D_{i}T_{i1}}$, $\rho_{D_{i}T_{i2}}$, and $\rho_{T_{i1}T_{i2}}$:

\begin{flalign*}
\rho_{D_{i}T_{i1}} &= \frac{(-1){{\sigma_{U_{i}}}^2}\psi\chi+(-1){{\sigma_{T_{i1}}}^2}\delta+{{\sigma_{T_{i1}}}^2}\theta+{{\sigma_{T_{i1}}}^2}\delta\phi+{{\sigma_{U_{i}}}^2}\delta\chi\gamma+{{\sigma_{U_{i}}}^2}\psi\chi}{\sigma_{D_{i}}\sigma_{T_{i1}}} \\ 
	&= \frac{-\delta{{\sigma_{T_{i1}}}^2}+{{\sigma_{T_{i1}}}^2}\theta+{{\sigma_{T_{i1}}}^2}\delta\phi+{{\sigma_{U_{i}}}^2}\delta\chi\gamma}{\sigma_{D_{i}}\sigma_{T_{i1}}}
\end{flalign*}
\begin{flalign*}
\rho_{D_{i}T_{i2}} &= \frac{\splitfrac{(-1){{\sigma_{U_{i}}}^2}\psi\gamma+(-1){{\sigma_{U_{i}}}^2}\delta\chi\phi+(-1){{\sigma_{U_{i}}}^2}\delta\chi\gamma+(-1){{\sigma_{T_{i1}}}^2}\delta\phi}{+{{\sigma_{U_{i}}}^2}\theta\chi\gamma+{{\sigma_{T_{i1}}}^2}\theta\phi+{{\sigma_{T_{i2}}}^2}\delta+{{\sigma_{U_{i}}}^2}\psi\chi\phi+{{\sigma_{U_{i}}}^2}\psi\gamma}}{\sigma_{D_{i}}\sigma_{T_{i2}}} \\ 
    &= \frac{-{{\sigma_{U_{i}}}^2}\delta\chi\gamma-{{\sigma_{T_{i1}}}^2}\delta\phi+{{\sigma_{U_{i}}}^2}\theta\chi\gamma+{{\sigma_{T_{i1}}}^2}\theta\phi+{{\sigma_{T_{i2}}}^2}\delta}{\sigma_{D_{i}}\sigma_{T_{i2}}}
\end{flalign*}
\begin{flalign*}
\rho_{T_{i1}T_{i2}} &= \frac{{{\sigma_{U_{i}}}^2}\chi\gamma+{{\sigma_{T_{i1}}}^2}\phi}{\sigma_{T_{i1}}\sigma_{T_{i2}}}
\end{flalign*}

\noindent We compute $\beta_{D_{i}T_{i1} \cdot T_{i2}}$ and $\beta_{D_{i}T_{i2} \cdot T_{i1}}$:

\begin{flalign*}
\beta_{D_{i}T_{i1} \cdot T_{i2}} &= 
\frac{\splitfrac{
{\frac{-\delta{{\sigma_{T_{i1}}}^2}+{{\sigma_{T_{i1}}}^2}\theta+{{\sigma_{T_{i1}}}^2}\delta\phi+{{\sigma_{U_{i}}}^2}\delta\chi\gamma}{\sigma_{D_{i}}\sigma_{T_{i1}}}}}
{-
{\frac{-{{\sigma_{U_{i}}}^2}\delta\chi\gamma-{{\sigma_{T_{i1}}}^2}\delta\phi+{{\sigma_{U_{i}}}^2}\theta\chi\gamma+{{\sigma_{T_{i1}}}^2}\theta\phi+{{\sigma_{T_{i2}}}^2}\delta}{\sigma_{D_{i}}\sigma_{T_{i2}}}}
*
{\frac{{{\sigma_{U_{i}}}^2}\chi\gamma+{{\sigma_{T_{i1}}}^2}\phi}{\sigma_{T_{i1}}\sigma_{T_{i2}}}}
}}
{
1-({\frac{{{\sigma_{U_{i}}}^2}\chi\gamma+{{\sigma_{T_{i1}}}^2}\phi}{\sigma_{T_{i1}}\sigma_{T_{i2}}}})^2
} 
*
\frac{\sigma_{D_{i}}}{\sigma_{T_{i1}}}\\ 
    &= \theta-\delta
\end{flalign*}

\begin{flalign*}
\beta_{D_{i}T_{i2} \cdot T_{i1}} &= \frac{\splitfrac{
{\frac{-{{\sigma_{U_{i}}}^2}\delta\chi\gamma-{{\sigma_{T_{i1}}}^2}\delta\phi+{{\sigma_{U_{i}}}^2}\theta\chi\gamma+{{\sigma_{T_{i1}}}^2}\theta\phi+{{\sigma_{T_{i2}}}^2}\delta}{\sigma_{D_{i}}\sigma_{T_{i2}}}}}
{-
{\frac{-\delta{{\sigma_{T_{i1}}}^2}+{{\sigma_{T_{i1}}}^2}\theta+{{\sigma_{T_{i1}}}^2}\delta\phi+{{\sigma_{U_{i}}}^2}\delta\chi\gamma}{\sigma_{D_{i}}\sigma_{T_{i1}}}}
*
{\frac{{{\sigma_{U_{i}}}^2}\chi\gamma+{{\sigma_{T_{i1}}}^2}\phi}{\sigma_{T_{i1}}\sigma_{T_{i2}}}}
}}
{
1-({\frac{{{\sigma_{U_{i}}}^2}\chi\gamma+{{\sigma_{T_{i1}}}^2}\phi}{\sigma_{T_{i1}}\sigma_{T_{i2}}}})^2
} 
*
\frac{\sigma_{D_{i}}}{\sigma_{T_{i2}}}\\ 
	&= \delta
\end{flalign*}

\noindent Finally, we compute $\textit{SC}$:
\begin{flalign*}
SC &= \beta_{D_{i}T_{i1} \cdot T_{i2}}+\beta_{D_{i}T_{i2} \cdot T_{i1}} \\
	&= \theta-\delta+\delta \\
	&= \theta \\
\end{flalign*}

\newpage


\section{Derivations for Models with Two-Sided Spillover}

\subsection*{\textit{Base model with two-sided spillover (Figure 2A)}}

We now consider models with two-sided exposure-to-outcome spillover ($T_{i1} \rightarrow Y_{i2}$ and $T_{i2} \rightarrow Y_{i1}$). Under the data generating process in \textbf{Figure 2A}, we can identify the difference between $\theta$ and $\kappa$. This may be informative depending on the value of $\kappa$ (for example, if $\kappa>0$, then $SC$ underestimates $\theta$). We compute $\rho_{D_{i}T_{i1}}$, $\rho_{D_{i}T_{i2}}$, and $\rho_{T_{i1}T_{i2}}$:

\begin{flalign*}
\rho_{D_{i}T_{i1}} &= \frac{(-1){\sigma_{U_{i}}}^2\psi\chi+(-1){\sigma_{T_{i1}}}^2\delta+(-1){\sigma_{U_{i}}}^2\kappa\chi\gamma+{\sigma_{T_{i1}}}^2\theta+{\sigma_{U_{i}}}^2\delta\chi\gamma+{\sigma_{U_{i}}}^2\psi\chi}{{\sigma_{D_{i}}\sigma_{T_{i1}}}} \\ 
	&= \frac{-{\sigma_{T_{i1}}}^2\delta-{\sigma_{U_{i}}}^2\kappa\chi\gamma+{\sigma_{T_{i1}}}^2\theta+{\sigma_{U_{i}}}^2\delta\chi\gamma}{{\sigma_{D_{i}}\sigma_{T_{i1}}}}
\end{flalign*}
\begin{flalign*}
\rho_{D_{i}T_{i2}} &= \frac{(-1){\sigma_{U_{i}}}^2\psi\gamma+(-1){\sigma_{U_{i}}}^2\delta\chi\gamma+(-1){\sigma_{T_{i2}}}^2\kappa+{\sigma_{U_{i}}}^2\theta\chi\gamma+{\sigma_{T_{i2}}}^2\delta+{\sigma_{U_{i}}}^2\psi\gamma}{{\sigma_{D_{i}}\sigma_{T_{i2}}}} \\ 
	&= \frac{-{\sigma_{U_{i}}}^2\delta\chi\gamma-{\sigma_{T_{i2}}}^2\kappa+{\sigma_{U_{i}}}^2\theta\chi\gamma+{\sigma_{T_{i2}}}^2\delta}{{\sigma_{D_{i}}\sigma_{T_{i2}}}}
\end{flalign*}
\begin{flalign*}
\rho_{T_{i1}T_{i2}} &= \frac{{{\sigma_{U_{i}}}^2}\chi\gamma}{\sigma_{T_{i1}}\sigma_{T_{i2}}} 
\end{flalign*}

\noindent We compute $\beta_{D_{i}T_{i1} \cdot T_{i2}}$ and $\beta_{D_{i}T_{i2} \cdot T_{i1}}$:

\begin{flalign*}
\beta_{D_{i}T_{i1} \cdot T_{i2}} &= \frac{
\frac{-{\sigma_{T_{i1}}}^2\delta-{\sigma_{U_{i}}}^2\kappa\chi\gamma+{\sigma_{T_{i1}}}^2\theta+{\sigma_{U_{i}}}^2\delta\chi\gamma}{{\sigma_{D_{i}}\sigma_{T_{i1}}}}
-
\frac{-{\sigma_{U_{i}}}^2\delta\chi\gamma-{\sigma_{T_{i2}}}^2\kappa+{\sigma_{U_{i}}}^2\theta\chi\gamma+{\sigma_{T_{i2}}}^2\delta}{{\sigma_{D_{i}}\sigma_{T_{i2}}}}
*
\frac{{{\sigma_{U_{i}}}^2}\chi\gamma}{\sigma_{T_{i1}}\sigma_{T_{i2}}} 
}
{
1 -
(\frac{{{\sigma_{U_{i}}}^2}\chi\gamma}{\sigma_{T_{i1}}\sigma_{T_{i2}}})^2
}
*
\frac{\sigma_{D_{i}}}{\sigma_{T_{i1}}} \\ 
    &= \theta-\delta
\end{flalign*}

\begin{flalign*}
\beta_{D_{i}T_{i2} \cdot T_{i1}} &= \frac{
\frac{-{\sigma_{U_{i}}}^2\delta\chi\gamma-{\sigma_{T_{i2}}}^2\kappa+{\sigma_{U_{i}}}^2\theta\chi\gamma+{\sigma_{T_{i2}}}^2\delta}{{\sigma_{D_{i}}\sigma_{T_{i2}}}}
-
\frac{-{\sigma_{T_{i1}}}^2\delta-{\sigma_{U_{i}}}^2\kappa\chi\gamma+{\sigma_{T_{i1}}}^2\theta+{\sigma_{U_{i}}}^2\delta\chi\gamma}{{\sigma_{D_{i}}\sigma_{T_{i1}}}}
*
\frac{{{\sigma_{U_{i}}}^2}\chi\gamma}{\sigma_{T_{i1}}\sigma_{T_{i2}}} 
}
{
1 -
(\frac{{{\sigma_{U_{i}}}^2}\chi\gamma}{\sigma_{T_{i1}}\sigma_{T_{i2}}})^2
}
*
\frac{\sigma_{D_{i}}}{\sigma_{T_{i2}}} \\ 
	&= \delta-\kappa
\end{flalign*}

\noindent Finally, we compute $\textit{SC}$:
\begin{flalign*}
SC   &= \beta_{D_{i}T_{i1} \cdot T_{i2}}+\beta_{D_{i}T_{i2} \cdot T_{i1}} \\
	&= \theta-\delta+\delta-\kappa \\
	&= \theta-\kappa \\
\end{flalign*}

\subsection*{\textit{Two-sided spillover and exposure-to-exposure spillover (Figures 2B-2C)}}

Under the data generating process in \textbf{Figure 2B}, we can still identify the difference between $\theta$ and $\kappa$ even with exposure-to-exposure spillover. We compute $\rho_{D_{i}T_{i1}}$, $\rho_{D_{i}T_{i2}}$, and $\rho_{T_{i1}T_{i2}}$:

\begin{flalign*}
\rho_{D_{i}T_{i1}} &=  \frac{\splitfrac{(-1){\sigma_{U_{i}}}^2\psi\chi+(-1){\sigma_{U_{i}}}^2\psi\tau\gamma+(-1){\sigma_{T_{i1}}}^2\delta+(-1){\sigma_{T_{i2}}}^2\kappa\tau+(-1){\sigma_{U_{i}}}^2\kappa\chi\gamma}{+{\sigma_{U_{i}}}^2\theta+{\sigma_{T_{i2}}}^2\delta\tau+{\sigma_{U_{i}}}^2\delta\chi\gamma+{\sigma_{U_{i}}}^2\psi\tau\gamma+{\sigma_{U_{i}}}^2\psi\chi}}{\sigma_{D_{i}}\sigma_{T_{i1}}} \\
	&= \frac{-{\sigma_{T_{i1}}}^2\delta-{\sigma_{T_{i2}}}^2\kappa\tau-{\sigma_{U_{i}}}^2\kappa\chi\gamma+{\sigma_{U_{i}}}^2\theta+{\sigma_{T_{i2}}}^2\delta\tau+{\sigma_{U_{i}}}^2\delta\chi\gamma}{\sigma_{D_{i}}\sigma_{T_{i1}}}
\end{flalign*}
\begin{flalign*}
\rho_{D_{i}T_{i2}} &=  \frac{\splitfrac{(-1){\sigma_{U_{i}}}^2\psi\gamma+(-1){\sigma_{U_{i}}}^2\delta\chi\gamma+(-1){\sigma_{T_{i2}}}^2\delta\tau+(-1){\sigma_{T_{i2}}}^2\kappa}{+{\sigma_{U_{i}}}^2\theta\chi\gamma+{\sigma_{T_{i2}}}^2\theta\tau+{\sigma_{T_{i2}}}^2\delta+{\sigma_{U_{i}}}^2\psi\gamma}}{\sigma_{D_{i}}\sigma_{T_{i2}}} \\ 
	&= \frac{-{\sigma_{U_{i}}}^2\delta\chi\gamma-{\sigma_{T_{i2}}}^2\delta\tau-{\sigma_{T_{i2}}}^2\kappa+{\sigma_{U_{i}}}^2\theta\chi\gamma+{\sigma_{T_{i2}}}^2\theta\tau+{\sigma_{T_{i2}}}^2\delta}{\sigma_{D_{i}}\sigma_{T_{i2}}}
\end{flalign*}
\begin{flalign*}
\rho_{T_{i1}T_{i2}} &= \frac{{{\sigma_{U_{i}}}^2}\chi\gamma+{{\sigma_{T_{i2}}}^2}\tau}{\sigma_{T_{i1}}\sigma_{T_{i2}}}
\end{flalign*}

\noindent We compute $\beta_{D_{i}T_{i1} \cdot T_{i2}}$ and $\beta_{D_{i}T_{i2} \cdot T_{i1}}$:

\begin{flalign*}
\beta_{D_{i}T_{i1} \cdot T_{i2}} &= \frac{\splitfrac{
\frac{-{\sigma_{T_{i1}}}^2\delta-{\sigma_{T_{i2}}}^2\kappa\tau-{\sigma_{U_{i}}}^2\kappa\chi\gamma+{\sigma_{U_{i}}}^2\theta+{\sigma_{T_{i2}}}^2\delta\tau+{\sigma_{U_{i}}}^2\delta\chi\gamma}{\sigma_{D_{i}}\sigma_{T_{i1}}}}
{
-
\frac{-{\sigma_{U_{i}}}^2\delta\chi\gamma-{\sigma_{T_{i2}}}^2\delta\tau-{\sigma_{T_{i2}}}^2\kappa+{\sigma_{U_{i}}}^2\theta\chi\gamma+{\sigma_{T_{i2}}}^2\theta\tau+{\sigma_{T_{i2}}}^2\delta}{\sigma_{D_{i}}\sigma_{T_{i2}}}
*
\frac{{{\sigma_{U_{i}}}^2}\chi\gamma+{{\sigma_{T_{i2}}}^2}\tau}{\sigma_{T_{i1}}\sigma_{T_{i2}}}
}}
{
1 -
(\frac{{{\sigma_{U_{i}}}^2}\chi\gamma+{{\sigma_{T_{i2}}}^2}\tau}{\sigma_{T_{i1}}\sigma_{T_{i2}}})^2
}
*
\frac{\sigma_{D_{i}}}{\sigma_{T_{i1}}} \\ 
    &= \theta-\delta
\end{flalign*}

\begin{flalign*}
\beta_{D_{i}T_{i2} \cdot T_{i1}} &= \frac{\splitfrac{
\frac{-{\sigma_{U_{i}}}^2\delta\chi\gamma-{\sigma_{T_{i2}}}^2\delta\tau-{\sigma_{T_{i2}}}^2\kappa+{\sigma_{U_{i}}}^2\theta\chi\gamma+{\sigma_{T_{i2}}}^2\theta\tau+{\sigma_{T_{i2}}}^2\delta}{\sigma_{D_{i}}\sigma_{T_{i2}}}}
{-
\frac{-{\sigma_{T_{i1}}}^2\delta-{\sigma_{T_{i2}}}^2\kappa\tau-{\sigma_{U_{i}}}^2\kappa\chi\gamma+{\sigma_{U_{i}}}^2\theta+{\sigma_{T_{i2}}}^2\delta\tau+{\sigma_{U_{i}}}^2\delta\chi\gamma}{\sigma_{D_{i}}\sigma_{T_{i1}}}
*
\frac{{{\sigma_{U_{i}}}^2}\chi\gamma+{{\sigma_{T_{i2}}}^2}\tau}{\sigma_{T_{i1}}\sigma_{T_{i2}}}
}}
{
1 -
(\frac{{{\sigma_{U_{i}}}^2}\chi\gamma+{{\sigma_{T_{i2}}}^2}\tau}{\sigma_{T_{i1}}\sigma_{T_{i2}}})^2
}
*
\frac{\sigma_{D_{i}}}{\sigma_{T_{i2}}} \\ 
	&= \delta-\kappa
\end{flalign*}

\noindent Finally, we compute $\textit{SC}$:
\begin{flalign*}
SC   &= \beta_{D_{i}T_{i1} \cdot T_{i2}}+\beta_{D_{i}T_{i2} \cdot T_{i1}} \\
	&= \theta-\delta+\delta-\kappa \\
	&= \theta-\kappa 
\end{flalign*}

\noindent We also consider the data generating process in \textbf{Figure 2C}. This is identical to that of \textbf{Figure 2B} except $T_{i1} \rightarrow T_{i2}$ with an effect of $\phi$. Changing the direction of the exposure-to-exposure spillover does not affect the partial regression coefficients and the spillover coefficient. 

\newpage

\section{Derivations for Models with Spillovers from Outcomes}

\subsection*{\textit{Past outcome affects future exposure (Figure 3A)}}

We lastly consider models in which spillovers originate from outcomes. Under the data generating process in \textbf{Figure 3A}, we do not identify the spillover effect $\theta$. We compute $\rho_{D_{i}T_{i1}}$, $\rho_{D_{i}T_{i2}}$, and $\rho_{T_{i1}T_{i2}}$:

\begin{flalign*}
\rho_{D_{i}T_{i1}} &=  \frac{(-1){\sigma_{U_{i}}}^2\psi\chi+(-1){\sigma_{T_{i1}}}^2\delta+{\sigma_{T_{i1}}}^2\theta+{\sigma_{T_{i1}}}^2\delta^2\omega+{\sigma_{U_{i}}}^2\delta\chi\gamma+{\sigma_{U_{i}}}^2\psi\chi}{\sigma_{D_{i}}\sigma_{T_{i1}}} \\ 
	&= \frac{-{\sigma_{T_{i1}}}^2\delta+{\sigma_{T_{i1}}}^2\theta+{\sigma_{T_{i1}}}^2\delta^2\omega+{\sigma_{U_{i}}}^2\delta\chi\gamma}{\sigma_{D_{i}}\sigma_{T_{i1}}}
\end{flalign*}
\begin{flalign*}
\rho_{D_{i}T_{i2}} &=  \frac{\splitfrac{(-1){\sigma_{U_{i}}}^2\psi\gamma+(-1){\sigma_{U_{i}}}^2\delta\chi\gamma+(-1){\sigma_{Y_{i1}}}^2\omega}{+{\sigma_{U_{i}}}^2\theta\chi\gamma+{\sigma_{U_{i}}}^2\theta\omega\psi\chi+{\sigma_{T_{i2}}}^2\delta+{\sigma_{U_{i}}}^2\delta\omega\psi\chi+{\sigma_{U_{i}}}^2\omega\psi^2+{\sigma_{U_{i}}}^2\psi\gamma}}{\sigma_{D_{i}}\sigma_{T_{i2}}} \\ 
	&= \frac{-{\sigma_{U_{i}}}^2\delta\chi\gamma-{\sigma_{Y_{i1}}}^2\omega+{\sigma_{U_{i}}}^2\theta\chi\gamma+{\sigma_{U_{i}}}^2\theta\omega\psi\chi+{\sigma_{T_{i2}}}^2\delta+{\sigma_{U_{i}}}^2\delta\omega\psi\chi+{\sigma_{U_{i}}}^2\omega\psi^2}{\sigma_{D_{i}}\sigma_{T_{i2}}}
\end{flalign*}
\begin{flalign*}
\rho_{T_{i1}T_{i2}} &= \frac{{\sigma_{U_{i}}}^2\chi\gamma+{\sigma_{U_{i}}}^2\omega\psi\chi+{\sigma_{T_{i1}}}^2\delta\omega}{\sigma_{T_{i1}}\sigma_{T_{i2}}}
\end{flalign*}

\noindent We compute $\beta_{D_{i}T_{i1} \cdot T_{i2}}$ and $\beta_{D_{i}T_{i2} \cdot T_{i1}}$: 

\begin{flalign*}
\beta_{D_{i}T_{i1} \cdot T_{i2}} &= \frac{\splitfrac{
\frac{-{\sigma_{T_{i1}}}^2\delta+{\sigma_{T_{i1}}}^2\theta+{\sigma_{T_{i1}}}^2\delta^2\omega+{\sigma_{U_{i}}}^2\delta\chi\gamma}{\sigma_{D_{i}}\sigma_{T_{i1}}}}
{
-
\frac{\splitfrac{-{\sigma_{U_{i}}}^2\delta\chi\gamma-{\sigma_{Y_{i1}}}^2\omega+{\sigma_{U_{i}}}^2\theta\chi\gamma}{+{\sigma_{U_{i}}}^2\theta\omega\psi\chi+{\sigma_{T_{i2}}}^2\delta+{\sigma_{U_{i}}}^2\delta\omega\psi\chi+{\sigma_{U_{i}}}^2\omega\psi^2}}{\sigma_{D_{i}}\sigma_{T_{i2}}}
*
\frac{{\sigma_{U_{i}}}^2\chi\gamma+{\sigma_{U_{i}}}^2\omega\psi\chi+{\sigma_{T_{i1}}}^2\delta\omega}{\sigma_{T_{i1}}\sigma_{T_{i2}}}
}}
{
1 -
(\frac{{\sigma_{U_{i}}}^2\chi\gamma+{\sigma_{U_{i}}}^2\omega\psi\chi+{\sigma_{T_{i1}}}^2\delta\omega}{\sigma_{T_{i1}}\sigma_{T_{i2}}})^2
}
*
\frac{\sigma_{D_{i}}}{\sigma_{T_{i1}}} 
\end{flalign*}

\begin{flalign*}
\beta_{D_{i}T_{i2} \cdot T_{i1}} &= \frac{\splitfrac{
\frac{-{\sigma_{U_{i}}}^2\delta\chi\gamma-{\sigma_{Y_{i1}}}^2\omega+{\sigma_{U_{i}}}^2\theta\chi\gamma+{\sigma_{U_{i}}}^2\theta\omega\psi\chi+{\sigma_{T_{i2}}}^2\delta+{\sigma_{U_{i}}}^2\delta\omega\psi\chi+{\sigma_{U_{i}}}^2\omega\psi^2}{\sigma_{D_{i}}\sigma_{T_{i2}}}}
{-
\frac{-{\sigma_{T_{i1}}}^2\delta+{\sigma_{T_{i1}}}^2\theta+{\sigma_{T_{i1}}}^2\delta^2\omega+{\sigma_{U_{i}}}^2\delta\chi\gamma}{\sigma_{D_{i}}\sigma_{T_{i1}}}
*
\frac{{\sigma_{U_{i}}}^2\chi\gamma+{\sigma_{U_{i}}}^2\omega\psi\chi+{\sigma_{T_{i1}}}^2\delta\omega}{\sigma_{T_{i1}}\sigma_{T_{i2}}}
}}
{
1 -
(\frac{{\sigma_{U_{i}}}^2\chi\gamma+{\sigma_{U_{i}}}^2\omega\psi\chi+{\sigma_{T_{i1}}}^2\delta\omega}{\sigma_{T_{i1}}\sigma_{T_{i2}}})^2
}
*
\frac{\sigma_{D_{i}}}{\sigma_{T_{i2}}} 
\end{flalign*}

\noindent In this data generating process, simplifying the partial regression coefficients will does not isolate the targeted effect $\delta$ nor the spillover effect $\theta$. Additionally, computing $SC$ by summing the previously calculated values of $\beta_{D_{i}T_{i1} \cdot T_{i2}}$ and $\beta_{D_{i}T_{i2} \cdot T_{i1}}$ will not identify $\theta$. 

\subsection*{\textit{Outcome-to-outcome spillover (Figures 3B-3C)}}

Under the data generating process in \textbf{Figure 3B}, sibling 1's outcome affects sibling 2's outcome ($Y_{i1} \rightarrow Y_{i2}$), and we do not identify the spillover effect $\theta$. We compute $\rho_{D_{i}T_{i1}}$, $\rho_{D_{i}T_{i2}}$, and $\rho_{T_{i1}T_{i2}}$:

\begin{flalign*}
\rho_{D_{i}T_{i1}} &=  \frac{(-1){\sigma_{U_{i}}}^2\psi\chi+(-1){\sigma_{T_{i1}}}^2\delta+{\sigma_{U_{i}}}^2\eta\psi\chi+{\sigma_{T_{i1}}}^2\eta\delta+{\sigma_{T_{i1}}}^2\theta+{\sigma_{U_{i}}}^2\delta\chi\gamma+{\sigma_{U_{i}}}^2\psi\chi}{\sigma_{D_{i}}\sigma_{T_{i1}}} \\
	&= \frac{-{\sigma_{T_{i1}}}^2\delta+{\sigma_{U_{i}}}^2\eta\psi\chi+{\sigma_{T_{i1}}}^2\eta\delta+{\sigma_{T_{i1}}}^2\theta+{\sigma_{U_{i}}}^2\delta\chi\gamma}{\sigma_{D_{i}}\sigma_{T_{i1}}} 
\end{flalign*}
\begin{flalign*}
\rho_{D_{i}T_{i2}} &=  \frac{(-1){\sigma_{U_{i}}}^2\psi\gamma+(-1){\sigma_{U_{i}}}^2\delta\chi\gamma+{\sigma_{U_{i}}}^2\eta\psi\gamma+{\sigma_{T_{i1}}}^2\eta\delta+{\sigma_{U_{i}}}^2\theta\chi\gamma+{\sigma_{T_{i2}}}^2\delta+{\sigma_{U_{i}}}^2\psi\gamma}{\sigma_{D_{i}}\sigma_{T_{i2}}} \\
	&= \frac{-{\sigma_{U_{i}}}^2\delta\chi\gamma+{\sigma_{U_{i}}}^2\eta\psi\gamma+{\sigma_{T_{i1}}}^2\eta\delta+{\sigma_{U_{i}}}^2\theta\chi\gamma+{\sigma_{T_{i2}}}^2\delta}{\sigma_{D_{i}}\sigma_{T_{i2}}}
\end{flalign*}
\begin{flalign*}
\rho_{T_{i1}T_{i2}} &= \frac{{\sigma_{U_{i}}}^2\chi\gamma}{\sigma_{T_{i1}}\sigma_{T_{i2}}}
\end{flalign*}

\noindent We compute $\beta_{D_{i}T_{i1} \cdot T_{i2}}$ and $\beta_{D_{i}T_{i2} \cdot T_{i1}}$: 

\begin{flalign*}
\beta_{D_{i}T_{i1} \cdot T_{i2}} &= \frac{\splitfrac{
\frac{-{\sigma_{T_{i1}}}^2\delta+{\sigma_{U_{i}}}^2\eta\psi\chi+{\sigma_{T_{i1}}}^2\eta\delta+{\sigma_{T_{i1}}}^2\theta+{\sigma_{U_{i}}}^2\delta\chi\gamma}{\sigma_{D_{i}}\sigma_{T_{i1}}}} 
{-
\frac{-{\sigma_{U_{i}}}^2\delta\chi\gamma+{\sigma_{U_{i}}}^2\eta\psi\gamma+{\sigma_{T_{i1}}}^2\eta\delta+{\sigma_{U_{i}}}^2\theta\chi\gamma+{\sigma_{T_{i2}}}^2\delta}{\sigma_{D_{i}}\sigma_{T_{i2}}}
*
\frac{{\sigma_{U_{i}}}^2\chi\gamma}{\sigma_{T_{i1}}\sigma_{T_{i2}}}}
}
{
1 -
(\frac{{{\sigma_{U_{i}}}^2}\chi\gamma+{{\sigma_{T_{i1}}}^2}\phi}{\sigma_{T_{i1}}\sigma_{T_{i2}}})^2
}
*
\frac{\sigma_{D_{i}}}{\sigma_{T_{i1}}} 
\end{flalign*}

\begin{flalign*}
\beta_{D_{i}T_{i2} \cdot T_{i1}} &= \frac{\splitfrac{
\frac{-{\sigma_{U_{i}}}^2\delta\chi\gamma+{\sigma_{U_{i}}}^2\eta\psi\gamma+{\sigma_{T_{i1}}}^2\eta\delta+{\sigma_{U_{i}}}^2\theta\chi\gamma+{\sigma_{T_{i2}}}^2\delta}{\sigma_{D_{i}}\sigma_{T_{i2}}}}
{-
\frac{-{\sigma_{T_{i1}}}^2\delta+{\sigma_{U_{i}}}^2\eta\psi\chi+{\sigma_{T_{i1}}}^2\eta\delta+{\sigma_{T_{i1}}}^2\theta+{\sigma_{U_{i}}}^2\delta\chi\gamma}{\sigma_{D_{i}}\sigma_{T_{i1}}}
*
\frac{{\sigma_{U_{i}}}^2\chi\gamma}{\sigma_{T_{i1}}\sigma_{T_{i2}}}}
}
{
1 -
(\frac{{\sigma_{U_{i}}}^2\chi\gamma}{\sigma_{T_{i1}}\sigma_{T_{i2}}})^2
}
*
\frac{\sigma_{D_{i}}}{\sigma_{T_{i2}}} 
\end{flalign*}

\noindent In this data generating process, simplifying the partial regression coefficients will does not isolate the targeted effect $\delta$ nor the spillover effect $\theta$. Additionally, computing $SC$ by summing the previously calculated values of $\beta_{D_{i}T_{i1} \cdot T_{i2}}$ and $\beta_{D_{i}T_{i2} \cdot T_{i1}}$ will not identify $\theta$. 

\break

\noindent The data generating process in \textbf{Figure 3C} has outcome-to-outcome spillover in which sibling 2's outcome affects sibling 1's outcome ($Y_{i2} \rightarrow Y_{i1}$). Similarly, we do not identify the spillover effect $\theta$. We compute $\rho_{D_{i}T_{i1}}$, $\rho_{D_{i}T_{i2}}$, and $\rho_{T_{i1}T_{i2}}$:

\begin{flalign*}
\rho_{D_{i}T_{i1}} &=  \frac{(-1){\sigma_{U_{i}}}^2\psi\chi+(-1){\sigma_{T_{i1}}}^2\delta+(-1){\sigma_{T_{i1}}}^2\lambda\theta+(-1){\sigma_{U_{i}}}^2\lambda\delta\chi\gamma+{\sigma_{U_{i}}}^2\delta\chi\gamma+{\sigma_{U_{i}}}^2\psi\chi}{\sigma_{D_{i}}\sigma_{T_{i1}}} \\
	&= \frac{-{\sigma_{T_{i1}}}^2\delta-{\sigma_{T_{i1}}}^2\lambda\theta-{\sigma_{U_{i}}}^2\lambda\delta\chi\gamma+{\sigma_{U_{i}}}^2\delta\chi\gamma}{\sigma_{D_{i}}\sigma_{T_{i1}}} 
\end{flalign*}
\begin{flalign*}
\rho_{D_{i}T_{i2}} &=  \frac{\splitfrac{(-1){\sigma_{U_{i}}}^2\psi\gamma+(-1){\sigma_{U_{i}}}^2\delta\chi\gamma+(-1){\sigma_{U_{i}}}^2\lambda\theta\chi\gamma+(-1){\sigma_{T_{i1}}}^2\lambda\delta}{+{\sigma_{U_{i}}}^2\theta\chi\gamma+{\sigma_{T_{i2}}}^2\delta+{\sigma_{U_{i}}}^2\psi\gamma}}{\sigma_{D_{i}}\sigma_{T_{i2}}} \\
	&= \frac{-{\sigma_{U_{i}}}^2\delta\chi\gamma-{\sigma_{U_{i}}}^2\lambda\theta\chi\gamma-{\sigma_{T_{i1}}}^2\lambda\delta+{\sigma_{U_{i}}}^2\theta\chi\gamma+{\sigma_{T_{i2}}}^2\delta}{\sigma_{D_{i}}\sigma_{T_{i2}}}
\end{flalign*}
\begin{flalign*}
\rho_{T_{i1}T_{i2}} &= \frac{{\sigma_{U_{i}}}^2\chi\gamma}{\sigma_{T_{i1}}\sigma_{T_{i2}}}
\end{flalign*}

\noindent We compute $\beta_{D_{i}T_{i1} \cdot T_{i2}}$ and $\beta_{D_{i}T_{i2} \cdot T_{i1}}$: 

\begin{flalign*}
\beta_{D_{i}T_{i1} \cdot T_{i2}} &= \frac{\splitfrac{
\frac{-{\sigma_{T_{i1}}}^2\delta-{\sigma_{T_{i1}}}^2\lambda\theta-{\sigma_{U_{i}}}^2\lambda\delta\chi\gamma+{\sigma_{U_{i}}}^2\delta\chi\gamma}{\sigma_{D_{i}}\sigma_{T_{i1}}}} 
{-
\frac{-{\sigma_{U_{i}}}^2\delta\chi\gamma-{\sigma_{U_{i}}}^2\lambda\theta\chi\gamma-{\sigma_{T_{i1}}}^2\lambda\delta+{\sigma_{U_{i}}}^2\theta\chi\gamma+{\sigma_{T_{i2}}}^2\delta}{\sigma_{D_{i}}\sigma_{T_{i2}}}
*
\frac{{\sigma_{U_{i}}}^2\chi\gamma}{\sigma_{T_{i1}}\sigma_{T_{i2}}}}
}
{
1 -
(\frac{{{\sigma_{U_{i}}}^2}\chi\gamma+{{\sigma_{T_{i1}}}^2}\phi}{\sigma_{T_{i1}}\sigma_{T_{i2}}})^2
}
*
\frac{\sigma_{D_{i}}}{\sigma_{T_{i1}}} 
\end{flalign*}

\begin{flalign*}
\beta_{D_{i}T_{i2} \cdot T_{i1}} &= \frac{\splitfrac{
\frac{-{\sigma_{U_{i}}}^2\delta\chi\gamma-{\sigma_{U_{i}}}^2\lambda\theta\chi\gamma-{\sigma_{T_{i1}}}^2\lambda\delta+{\sigma_{U_{i}}}^2\theta\chi\gamma+{\sigma_{T_{i2}}}^2\delta}{\sigma_{D_{i}}\sigma_{T_{i2}}}}
{-
\frac{-{\sigma_{T_{i1}}}^2\delta-{\sigma_{T_{i1}}}^2\lambda\theta-{\sigma_{U_{i}}}^2\lambda\delta\chi\gamma+{\sigma_{U_{i}}}^2\delta\chi\gamma}{\sigma_{D_{i}}\sigma_{T_{i1}}}
*
\frac{{\sigma_{U_{i}}}^2\chi\gamma}{\sigma_{T_{i1}}\sigma_{T_{i2}}}}
}
{
1 -
(\frac{{\sigma_{U_{i}}}^2\chi\gamma}{\sigma_{T_{i1}}\sigma_{T_{i2}}})^2
}
*
\frac{\sigma_{D_{i}}}{\sigma_{T_{i2}}} 
\end{flalign*}

\noindent In this data generating process, simplifying the partial regression coefficients will does not isolate the targeted effect $\delta$ nor the spillover effect $\theta$. Additionally, computing $SC$ by summing the previously calculated values of $\beta_{D_{i}T_{i1} \cdot T_{i2}}$ and $\beta_{D_{i}T_{i2} \cdot T_{i1}}$ will not identify $\theta$. 

\newpage

\section{Sampling Description for Empirical Application}

For our application, we estimated the effect of a younger sibling's preterm birth (gestational age $<$37 completed weeks) on their older sibling's Phonological Awareness Literacy Screening-Kindergarten (PALS-K) test score. PALS-K evaluates fundamental literacy skills at kindergarten entry.\cite{ford2014, pktech} We used data from Big Data for Little Kids (BD4LK), a cohort of birth records for live resident in-state deliveries in Wisconsin during 2007-2016 (N$>$666,000 births). Birth records link to multiple administrative sources, including paid Medicaid claims and encounters (2007-2016) and children's PALS-K scores (2012-2016 school years). BD4LK's linkage process has been previously described.\cite{larson2019,mallinson2019}

We sampled sibling pairs born sequentially to the same mother between January 1, 2007-September 1, 2010 and took the English-language PALS-K test. We restricted eligibility on birthdate because children had to be at least five years old by September 1, 2015 for PALS-K testing eligibility.\cite{widpi} PALS-K is available in two languages: English and Spanish.\cite{ford2014, pktech} However, test versions were developed separately and are not directly comparable, so we only considered children who took the English-language test.

BD4LK includes 252,883 unique deliveries during January 1, 2007-September 1, 2010. We identified 806 records (0.3\%) with multiple maternal or child identifiers that imperfectly matched to Medicaid claims or PALS-K scores. Among those records, we randomly selected one match and excluded the remainder. We then identified 177,863 records (70.3\%) that linked to PALS-K scores, of which 46,743 records were in-sample siblings (26.3\% of test-linked records). Finally, we pulled eligible sibling pairs: sequentially-born from different deliveries, took the English-language PALS-K test, and completed information on control variables (maternal age, maternal education, and Medicaid delivery coverage, all of which were measured at the older sibling's delivery). This generated a final sample of 40,020 siblings (85.6\% of tested siblings), or 20,010 sibling pairs. 

\newpage


\section{Supplemental Tables}


\begin{table}[ht]
\setlength\extrarowheight{5pt}
\captionsetup{justification=justified, singlelinecheck=false}
\caption{\textbf{Supplemental Table 1.} Descriptive statistics of sibling pairs for the empirical application (N = 20,010 sibling pairs$^{a}$)}
\begin{adjustbox}{max height=\textheight}{%
\begin{tabular} { l r r }
\hline
{} & \textbf{Older Siblings} & \textbf{Younger Siblings} \\
{} & \textbf{(N = 20,010)} & \textbf{(N = 20,010)} \\
\hline
PALS-K Score (Points)$^{c}$, Mean (SD) & {63.58 (24.12)} & {64.22 (23.83)} \\
Preterm Birth$^{b}$, N (\%) & {1,357 (6.78\%)} & {1,331 (6.65\%)} \\
Maternal Age (Years)$^{d}$, Mean (SD) & {26.01 (5.12)} & {--} \\
Maternal Education$^{d}$, N (\%) & {} & {} \\
\textit{No high school degree} & {2,865 (14.32\%)} & {--} \\
\textit{High school degree/equivalent only} & {5,789 (28.93\%)} & {--} \\
\textit{1-3 years college} & {5,036 (25.17\%)} & {--} \\
\textit{4+ years college} & {6,320 (31.58\%)} & {--} \\
Medicaid-Paid Delivery$^{d}$, N (\%) & {7,528 (37.62\%)} & {--} \\
\hline
\end{tabular}
}
\end{adjustbox}
\setlength\extrarowheight{0pt}
\caption{$^{a}$Each sibling pair includes one older sibling and one younger sibling. The sample consists of  40,020 children in total.}
\caption{$^{b}$Preterm birth is defined as gestational age $<$37 completed weeks.} 
\caption{$^{c}$PALS-K has a score range of 0-102 points.} 
\caption{$^{d}$Measured at the older sibling's delivery.} 
\caption{Notes: "PALS-K" Phonological Awareness Literacy Screening-Kindergarten, "SD" standard deviation.}
\end{table}


\newpage


\begin{table}[ht]
\setlength\extrarowheight{5pt}
\captionsetup{justification=justified, singlelinecheck=false}
\caption{\textbf{Supplemental Table 2.} Cross-tabulation of preterm birth$^{a}$ by sibling within two-sibling clusters (N = 20,010 pairs$^{b}$)}
\begin{adjustbox}{max height=\textheight}{%
\begin{tabular} { l | r | r | r |}
{} & \textbf{Younger Sibling} & \textbf{Younger Sibling} & \textbf{Total} \\
{} & \textbf{Not Preterm} & \textbf{Preterm} & {} \\
\hline
\textbf{Older Sibling} & {17,652} & {1,001} & {18,653} \\
\textbf{Not Preterm} & {(94.63\%)} & {(5.37\%)} & {(100.00\%)} \\
{} & {(94.50\%)} & {(74.21\%)} & {(93.22\%)} \\
\hline
\textbf{Older Sibling} & {1,027} & {330} & {1,357} \\
\textbf{Preterm} & {(75.68\%)} & {(24.32\%)} & {(100.00\%)} \\
{} & {(5.50\%)} & {(24.79\%)} & {(6.78\%)} \\
\hline
\textbf{Total} & {18,679} & {1,331} & {20,010} \\
{} & {(93.35\%)} & {(6.65\%)} & {(100.00\%)} \\
{} & {(100.00\%)} & {(100.00\%)} & {(100.00\%)} \\
\hline
\end{tabular}
}
\end{adjustbox}
\setlength\extrarowheight{0pt}
\caption{$^{a}$Preterm birth is defined as gestational age $<$37 completed weeks.} 
\caption{$^{b}$Each sibling pair includes one older sibling and one younger sibling. The sample consists of  40,020 children in total.}
\caption{Notes: Whole numbers indicate the frequency of sibling pairs. Within a cell, the first percentage indicates the row percentage (i.e., the percent of younger siblings who are preterm or not preterm by the older sibling's preterm birth status), and the second percentage is a column percentage (i.e., the percent of older siblings who are preterm or not preterm by the younger sibling's preterm birth status).}
\end{table}


\newpage


\section{Simulation Code}

We conducted Monte Carlo simulations in Stata Statistical Software: Release 16.\cite{statasoft} 

\begin{singlespace}
\begin{lstlisting}[language=Stata,style=stata-editor]
/*
Estimatinng Sibling Spillovers with 
Unobserved Confounding Using Gain-Scores

Simulation Code

David C. Mallinson, Felix Elwert
*/

/**********************/
*Preliminaries
/**********************/

/*
Defintions:

t_1 = sibling 1's treatment
t_2 = sibling 2's treatment
y_1 = sibling 1's outcome
y_2 = sibling 2's outcome
u = unobserved fixed effect
d = gain-score (y_2-y_1)
*/


/**********************/
*Simulate Model 1A
/**********************/

**clear previous data
clear all
set more off

**monte carlo simulation settings 
global nobs=5000 /*observations per run*/
global nmc=1000 /*number of runs*/
set seed 54385 /*seed for random number generator -- can change*/
set obs $nobs /*sets observations within each run*/

**set parameters (causal effects in models)
scalar delta=1 /*t_1 on y_1, t_2 on y_2*/

scalar psi=1 /*u on y_1, u on y_2*/
scalar chi=2 /*u on t_1*/
scalar gamma=3 /*u on t_2*/

scalar theta=0.5 /*t_1 on y_2*/
scalar kappa=0.3 /*t_2 on y_1*/
scalar tau=0.3 /*t_1 on t_2, t_2 on t_1*/

scalar omega=0.3 /*y_1 on t_2*/
scalar eta=0.3 /*y_1 on y_2*/
scalar lambda=0.3 /*y_2 on y_1*/

**generate variables

gen u=. /*unobserved fixed effect*/
gen t_1=. /*sibling 1's treatment*/
gen t_2=. /*sibling 2's treatment*/
gen y_1=. /*sibling 1's outcome*/
gen y_2=. /*sibling 2's outcome*/
gen d=. /*gain-score (y_2-y_1)*/

**simulation 

tempname m1a
postfile `m1a' sc_mc1 sc_ub_mc1 sc_lb_mc1 using m1a, replace
quietly {

forvalues i=1/$nmc {

	/*note: for y_1 and y_2 equations, "rnormal(0,1)" is residual*/

	replace u=rnormal(0,1)
	replace t_1=(chi*u)>0.5
	replace t_2=(gamma*u)>0.2
	replace y_1=(delta*t_1)+(psi*u)+rnormal(0,1)
	replace y_2=(delta*t_2)+(theta*t_1)+(psi*u)+rnormal(0,1)
	replace d=y_2-y_1
	
	/*gain-score regression*/
	reg d t_1 t_2
	
	/*lincom to compute spillover coefficient*/
	lincom t_1+t_2
	
	/*store lincom output*/
	scalar sc_mc1=r(estimate) /*spillover coefficient*/
	scalar sc_ub_mc1=(r(estimate)+(1.96*r(se))) /*95% CI upper bound*/
	scalar sc_lb_mc1=(r(estimate)-(1.96*r(se))) /*95% CI lower bound*/
	
	post `m1a' (sc_mc1) (sc_ub_mc1) (sc_lb_mc1)
}
}
postclose `m1a'

use m1a, clear
summarize

**store mean of results, save to append with other simulation results
egen sc=mean(sc_mc1)
egen ub=mean(sc_ub_mc1)
egen lb=mean(sc_lb_mc1)
gen id=1 /*first simulation*/

keep sc ub lb id
duplicates drop
duplicates report 

**save data 
save "sim1.dta", replace


/**********************/
*Simulate Model 1B
/**********************/

**clear previous data
clear all
set more off

**monte carlo simulation settings 
global nobs=5000 /*observations per run*/
global nmc=1000 /*number of runs*/
set seed 2414830 /*seed for random number generator -- can change*/
set obs $nobs /*sets observations within each run*/

**set parameters (causal effects in models)
scalar delta=1 /*t_1 on y_1, t_2 on y_2*/

scalar psi=1 /*u on y_1, u on y_2*/
scalar chi=2 /*u on t_1*/
scalar gamma=3 /*u on t_2*/

scalar theta=0.5 /*t_1 on y_2*/
scalar kappa=0.3 /*t_2 on y_1*/
scalar tau=0.3 /*t_1 on t_2, t_2 on t_1*/

scalar omega=0.3 /*y_1 on t_2*/
scalar eta=0.3 /*y_1 on y_2*/
scalar lambda=0.3 /*y_2 on y_1*/

**generate variables

gen u=. /*unobserved fixed effect*/
gen t_1=. /*sibling 1's treatment*/
gen t_2=. /*sibling 2's treatment*/
gen y_1=. /*sibling 1's outcome*/
gen y_2=. /*sibling 2's outcome*/
gen d=. /*gain-score (y_2-y_1)*/

**simulation 

tempname m1b
postfile `m1b' sc_mc2 sc_ub_mc2 sc_lb_mc2 using m1b, replace
quietly {

forvalues i=1/$nmc {

	/*note: for y_1 and y_2 equations, "rnormal(0,1)" is residual*/

	replace u=rnormal(0,1)
	replace t_1=(chi*u)>0.5
	replace t_2=(tau*t_1)+(gamma*u)>0.2
	replace y_1=(delta*t_1)+(psi*u)+rnormal(0,1)
	replace y_2=(delta*t_2)+(theta*t_1)+(psi*u)+rnormal(0,1)
	replace d=y_2-y_1
	
	/*gain-score regression*/
	reg d t_1 t_2
	
	/*lincom to compute spillover coefficient*/
	lincom t_1+t_2
	
	/*store lincom output*/
	scalar sc_mc2=r(estimate) /*spillover coefficient*/
	scalar sc_ub_mc2=(r(estimate)+(1.96*r(se))) /*95% CI upper bound*/
	scalar sc_lb_mc2=(r(estimate)-(1.96*r(se))) /*95% CI lower bound*/
	
	post `m1b' (sc_mc2) (sc_ub_mc2) (sc_lb_mc2)
}
}
postclose `m1b'

use m1b, clear
summarize

**store mean of results, save to append with other simulation results
egen sc=mean(sc_mc2)
egen ub=mean(sc_ub_mc2)
egen lb=mean(sc_lb_mc2)
gen id=2 /*second simulation*/

keep sc ub lb id
duplicates drop
duplicates report 

**save data 
save "sim2.dta", replace


/**********************/
*Simulate Model 1C
/**********************/

**clear previous data
clear all
set more off

**monte carlo simulation settings 
global nobs=5000 /*observations per run*/
global nmc=1000 /*number of runs*/
set seed 48910 /*seed for random number generator -- can change*/
set obs $nobs /*sets observations within each run*/

**set parameters (causal effects in models)
scalar delta=1 /*t_1 on y_1, t_2 on y_2*/

scalar psi=1 /*u on y_1, u on y_2*/
scalar chi=2 /*u on t_1*/
scalar gamma=3 /*u on t_2*/

scalar theta=0.5 /*t_1 on y_2*/
scalar kappa=0.3 /*t_2 on y_1*/
scalar tau=0.3 /*t_1 on t_2, t_2 on t_1*/

scalar omega=0.3 /*y_1 on t_2*/
scalar eta=0.3 /*y_1 on y_2*/
scalar lambda=0.3 /*y_2 on y_1*/

**generate variables

gen u=. /*unobserved fixed effect*/
gen t_1=. /*sibling 1's treatment*/
gen t_2=. /*sibling 2's treatment*/
gen y_1=. /*sibling 1's outcome*/
gen y_2=. /*sibling 2's outcome*/
gen d=. /*gain-score (y_2-y_1)*/

**simulation 

tempname m1c
postfile `m1c' sc_mc3 sc_ub_mc3 sc_lb_mc3 using m1c, replace
quietly {

forvalues i=1/$nmc {

	/*note: for y_1 and y_2 equations, "rnormal(0,1)" is residual*/

	replace u=rnormal(0,1)
	replace t_2=(gamma*u)>0.2
	replace t_1=(tau*t_2)+(chi*u)>0.5
	replace y_1=(delta*t_1)+(psi*u)+rnormal(0,1)
	replace y_2=(delta*t_2)+(theta*t_1)+(psi*u)+rnormal(0,1)
	replace d=y_2-y_1
	
	/*gain-score regression*/
	reg d t_1 t_2
	
	/*lincom to compute spillover coefficient*/
	lincom t_1+t_2
	
	/*store lincom output*/
	scalar sc_mc3=r(estimate) /*spillover coefficient*/
	scalar sc_ub_mc3=(r(estimate)+(1.96*r(se))) /*95% CI upper bound*/
	scalar sc_lb_mc3=(r(estimate)-(1.96*r(se))) /*95% CI lower bound*/
	
	post `m1c' (sc_mc3) (sc_ub_mc3) (sc_lb_mc3)
}
}
postclose `m1c'

use m1c, clear
summarize

**store mean of results, save to append with other simulation results
egen sc=mean(sc_mc3)
egen ub=mean(sc_ub_mc3)
egen lb=mean(sc_lb_mc3)
gen id=3 /*third simulation*/

keep sc ub lb id
duplicates drop
duplicates report 

**save data 
save "sim3.dta", replace


/**********************/
*Simulate Model 2A
/**********************/

**clear previous data
clear all
set more off

**monte carlo simulation settings 
global nobs=5000 /*observations per run*/
global nmc=1000 /*number of runs*/
set seed 9842 /*seed for random number generator -- can change*/
set obs $nobs /*sets observations within each run*/

**set parameters (causal effects in models)
scalar delta=1 /*t_1 on y_1, t_2 on y_2*/

scalar psi=1 /*u on y_1, u on y_2*/
scalar chi=2 /*u on t_1*/
scalar gamma=3 /*u on t_2*/

scalar theta=0.5 /*t_1 on y_2*/
scalar kappa=0.3 /*t_2 on y_1*/
scalar tau=0.3 /*t_1 on t_2, t_2 on t_1*/

scalar omega=0.3 /*y_1 on t_2*/
scalar eta=0.3 /*y_1 on y_2*/
scalar lambda=0.3 /*y_2 on y_1*/

**generate variables

gen u=. /*unobserved fixed effect*/
gen t_1=. /*sibling 1's treatment*/
gen t_2=. /*sibling 2's treatment*/
gen y_1=. /*sibling 1's outcome*/
gen y_2=. /*sibling 2's outcome*/
gen d=. /*gain-score (y_2-y_1)*/

**simulation 

tempname m2a
postfile `m2a' sc_mc4 sc_ub_mc4 sc_lb_mc4 using m2a, replace
quietly {

forvalues i=1/$nmc {

	/*note: for y_1 and y_2 equations, "rnormal(0,1)" is residual*/

	replace u=rnormal(0,1)
	replace t_1=(chi*u)>0.5
	replace t_2=(gamma*u)>0.2
	replace y_1=(delta*t_1)+(kappa*t_2)+(psi*u)+rnormal(0,1)
	replace y_2=(delta*t_2)+(theta*t_1)+(psi*u)+rnormal(0,1)
	replace d=y_2-y_1
	
	/*gain-score regression*/
	reg d t_1 t_2
	
	/*lincom to compute spillover coefficient*/
	lincom t_1+t_2
	
	/*store lincom output*/
	scalar sc_mc4=r(estimate) /*spillover coefficient*/
	scalar sc_ub_mc4=(r(estimate)+(1.96*r(se))) /*95% CI upper bound*/
	scalar sc_lb_mc4=(r(estimate)-(1.96*r(se))) /*95% CI lower bound*/
	
	post `m2a' (sc_mc4) (sc_ub_mc4) (sc_lb_mc4)
}
}
postclose `m2a'

use m2a, clear
summarize

**store mean of results, save to append with other simulation results
egen sc=mean(sc_mc4)
egen ub=mean(sc_ub_mc4)
egen lb=mean(sc_lb_mc4)
gen id=4 /*fourth simulation*/

keep sc ub lb id
duplicates drop
duplicates report 

**save data 
save "sim4.dta", replace


/**********************/
*Simulate Model 2B
/**********************/

**clear previous data
clear all
set more off

**monte carlo simulation settings 
global nobs=5000 /*observations per run*/
global nmc=1000 /*number of runs*/
set seed 33305 /*seed for random number generator -- can change*/
set obs $nobs /*sets observations within each run*/

**set parameters (causal effects in models)
scalar delta=1 /*t_1 on y_1, t_2 on y_2*/

scalar psi=1 /*u on y_1, u on y_2*/
scalar chi=2 /*u on t_1*/
scalar gamma=3 /*u on t_2*/

scalar theta=0.5 /*t_1 on y_2*/
scalar kappa=0.3 /*t_2 on y_1*/
scalar tau=0.3 /*t_1 on t_2, t_2 on t_1*/

scalar omega=0.3 /*y_1 on t_2*/
scalar eta=0.3 /*y_1 on y_2*/
scalar lambda=0.3 /*y_2 on y_1*/

**generate variables

gen u=. /*unobserved fixed effect*/
gen t_1=. /*sibling 1's treatment*/
gen t_2=. /*sibling 2's treatment*/
gen y_1=. /*sibling 1's outcome*/
gen y_2=. /*sibling 2's outcome*/
gen d=. /*gain-score (y_2-y_1)*/

**simulation 

tempname m2b
postfile `m2b' sc_mc5 sc_ub_mc5 sc_lb_mc5 using m2b, replace
quietly {

forvalues i=1/$nmc {

	/*note: for y_1 and y_2 equations, "rnormal(0,1)" is residual*/

	replace u=rnormal(0,1)
	replace t_1=(chi*u)>0.5
	replace t_2=(tau*t_1)+(gamma*u)>0.2
	replace y_1=(delta*t_1)+(kappa*t_2)+(psi*u)+rnormal(0,1)
	replace y_2=(delta*t_2)+(theta*t_1)+(psi*u)+rnormal(0,1)
	replace d=y_2-y_1
	
	/*gain-score regression*/
	reg d t_1 t_2
	
	/*lincom to compute spillover coefficient*/
	lincom t_1+t_2
	
	/*store lincom output*/
	scalar sc_mc5=r(estimate) /*spillover coefficient*/
	scalar sc_ub_mc5=(r(estimate)+(1.96*r(se))) /*95% CI upper bound*/
	scalar sc_lb_mc5=(r(estimate)-(1.96*r(se))) /*95% CI lower bound*/
	
	post `m2b' (sc_mc5) (sc_ub_mc5) (sc_lb_mc5)
}
}
postclose `m2b'

use m2b, clear
summarize

**store mean of results, save to append with other simulation results
egen sc=mean(sc_mc5)
egen ub=mean(sc_ub_mc5)
egen lb=mean(sc_lb_mc5)
gen id=5 /*fifth simulation*/

keep sc ub lb id
duplicates drop
duplicates report 

**save data 
save "sim5.dta", replace


/**********************/
*Simulate Model 2C
/**********************/

**clear previous data
clear all
set more off

**monte carlo simulation settings 
global nobs=5000 /*observations per run*/
global nmc=1000 /*number of runs*/
set seed 101678 /*seed for random number generator -- can change*/
set obs $nobs /*sets observations within each run*/

**set parameters (causal effects in models)
scalar delta=1 /*t_1 on y_1, t_2 on y_2*/

scalar psi=1 /*u on y_1, u on y_2*/
scalar chi=2 /*u on t_1*/
scalar gamma=3 /*u on t_2*/

scalar theta=0.5 /*t_1 on y_2*/
scalar kappa=0.3 /*t_2 on y_1*/
scalar tau=0.3 /*t_1 on t_2, t_2 on t_1*/

scalar omega=0.3 /*y_1 on t_2*/
scalar eta=0.3 /*y_1 on y_2*/
scalar lambda=0.3 /*y_2 on y_1*/

**generate variables

gen u=. /*unobserved fixed effect*/
gen t_1=. /*sibling 1's treatment*/
gen t_2=. /*sibling 2's treatment*/
gen y_1=. /*sibling 1's outcome*/
gen y_2=. /*sibling 2's outcome*/
gen d=. /*gain-score (y_2-y_1)*/

**simulation 

tempname m2c
postfile `m2c' sc_mc6 sc_ub_mc6 sc_lb_mc6 using m2c, replace
quietly {

forvalues i=1/$nmc {

	/*note: for y_1 and y_2 equations, "rnormal(0,1)" is residual*/

	replace u=rnormal(0,1)
	replace t_2=(gamma*u)>0.2
	replace t_1=(tau*t_2)+(chi*u)>0.5
	replace y_1=(delta*t_1)+(kappa*t_2)+(psi*u)+rnormal(0,1)
	replace y_2=(delta*t_2)+(theta*t_1)+(psi*u)+rnormal(0,1)
	replace d=y_2-y_1
	
	/*gain-score regression*/
	reg d t_1 t_2
	
	/*lincom to compute spillover coefficient*/
	lincom t_1+t_2
	
	/*store lincom output*/
	scalar sc_mc6=r(estimate) /*spillover coefficient*/
	scalar sc_ub_mc6=(r(estimate)+(1.96*r(se))) /*95% CI upper bound*/
	scalar sc_lb_mc6=(r(estimate)-(1.96*r(se))) /*95% CI lower bound*/
	
	post `m2c' (sc_mc6) (sc_ub_mc6) (sc_lb_mc6)
}
}
postclose `m2c'

use m2c, clear
summarize

**store mean of results, save to append with other simulation results
egen sc=mean(sc_mc6)
egen ub=mean(sc_ub_mc6)
egen lb=mean(sc_lb_mc6)
gen id=6 /*sixth simulation*/

keep sc ub lb id
duplicates drop
duplicates report 

**save data 
save "sim6.dta", replace


/**********************/
*Simulate Model 3A
/**********************/

**clear previous data
clear all
set more off

**monte carlo simulation settings 
global nobs=5000 /*observations per run*/
global nmc=1000 /*number of runs*/
set seed 278 /*seed for random number generator -- can change*/
set obs $nobs /*sets observations within each run*/

**set parameters (causal effects in models)
scalar delta=1 /*t_1 on y_1, t_2 on y_2*/

scalar psi=1 /*u on y_1, u on y_2*/
scalar chi=2 /*u on t_1*/
scalar gamma=3 /*u on t_2*/

scalar theta=0.5 /*t_1 on y_2*/
scalar kappa=0.3 /*t_2 on y_1*/
scalar tau=0.3 /*t_1 on t_2, t_2 on t_1*/

scalar omega=0.3 /*y_1 on t_2*/
scalar eta=0.3 /*y_1 on y_2*/
scalar lambda=0.3 /*y_2 on y_1*/

**generate variables

gen u=. /*unobserved fixed effect*/
gen t_1=. /*sibling 1's treatment*/
gen t_2=. /*sibling 2's treatment*/
gen y_1=. /*sibling 1's outcome*/
gen y_2=. /*sibling 2's outcome*/
gen d=. /*gain-score (y_2-y_1)*/

**simulation 

tempname m3a
postfile `m3a' sc_mc7 sc_ub_mc7 sc_lb_mc7 using m3a, replace
quietly {

forvalues i=1/$nmc {

	/*note: for y_1 and y_2 equations, "rnormal(0,1)" is residual*/

	replace u=rnormal(0,1)
	replace t_1=(chi*u)>0.5
	replace y_1=(delta*t_1)+(psi*u)+rnormal(0,1)
	replace t_2=(omega*y_1)+(gamma*u)>0.2
	replace y_2=(delta*t_2)+(theta*t_1)+(psi*u)+rnormal(0,1)
	replace d=y_2-y_1
	
	/*gain-score regression*/
	reg d t_1 t_2
	
	/*lincom to compute spillover coefficient*/
	lincom t_1+t_2
	
	/*store lincom output*/
	scalar sc_mc7=r(estimate) /*spillover coefficient*/
	scalar sc_ub_mc7=(r(estimate)+(1.96*r(se))) /*95% CI upper bound*/
	scalar sc_lb_mc7=(r(estimate)-(1.96*r(se))) /*95% CI lower bound*/
	
	post `m3a' (sc_mc7) (sc_ub_mc7) (sc_lb_mc7)
}
}
postclose `m3a'

use m3a, clear
summarize

**store mean of results, save to append with other simulation results
egen sc=mean(sc_mc7)
egen ub=mean(sc_ub_mc7)
egen lb=mean(sc_lb_mc7)
gen id=7 /*seventh simulation*/

keep sc ub lb id
duplicates drop
duplicates report 

**save data 
save "sim7.dta", replace


/**********************/
*Simulate Model 3B
/**********************/

**clear previous data
clear all
set more off

**monte carlo simulation settings 
global nobs=5000 /*observations per run*/
global nmc=1000 /*number of runs*/
set seed 53893 /*seed for random number generator -- can change*/
set obs $nobs /*sets observations within each run*/

**set parameters (causal effects in models)
scalar delta=1 /*t_1 on y_1, t_2 on y_2*/

scalar psi=1 /*u on y_1, u on y_2*/
scalar chi=2 /*u on t_1*/
scalar gamma=3 /*u on t_2*/

scalar theta=0.5 /*t_1 on y_2*/
scalar kappa=0.3 /*t_2 on y_1*/
scalar tau=0.3 /*t_1 on t_2, t_2 on t_1*/

scalar omega=0.3 /*y_1 on t_2*/
scalar eta=0.3 /*y_1 on y_2*/
scalar lambda=0.3 /*y_2 on y_1*/

**generate variables

gen u=. /*unobserved fixed effect*/
gen t_1=. /*sibling 1's treatment*/
gen t_2=. /*sibling 2's treatment*/
gen y_1=. /*sibling 1's outcome*/
gen y_2=. /*sibling 2's outcome*/
gen d=. /*gain-score (y_2-y_1)*/

**simulation 

tempname m3b
postfile `m3b' sc_mc8 sc_ub_mc8 sc_lb_mc8 using m3b, replace
quietly {

forvalues i=1/$nmc {

	/*note: for y_1 and y_2 equations, "rnormal(0,1)" is residual*/

	replace u=rnormal(0,1)
	replace t_1=(chi*u)>0.5
	replace t_2=(gamma*u)>0.2
	replace y_1=(delta*t_1)+(psi*u)+rnormal(0,1)
	replace y_2=(delta*t_2)+(theta*t_1)+(eta*y_1)+(psi*u)+rnormal(0,1)
	replace d=y_2-y_1
	
	/*gain-score regression*/
	reg d t_1 t_2
	
	/*lincom to compute spillover coefficient*/
	lincom t_1+t_2
	
	/*store lincom output*/
	scalar sc_mc8=r(estimate) /*spillover coefficient*/
	scalar sc_ub_mc8=(r(estimate)+(1.96*r(se))) /*95% CI upper bound*/
	scalar sc_lb_mc8=(r(estimate)-(1.96*r(se))) /*95% CI lower bound*/
	
	post `m3b' (sc_mc8) (sc_ub_mc8) (sc_lb_mc8)
}
}
postclose `m3b'

use m3b, clear
summarize

**store mean of results, save to append with other simulation results
egen sc=mean(sc_mc8)
egen ub=mean(sc_ub_mc8)
egen lb=mean(sc_lb_mc8)
gen id=8 /*eighth simulation*/

keep sc ub lb id
duplicates drop
duplicates report 

**save data 
save "sim8.dta", replace


/**********************/
*Simulate Model 3C
/**********************/

**clear previous data
clear all
set more off

**monte carlo simulation settings 
global nobs=5000 /*observations per run*/
global nmc=1000 /*number of runs*/
set seed 9921 /*seed for random number generator -- can change*/
set obs $nobs /*sets observations within each run*/

**set parameters (causal effects in models)
scalar delta=1 /*t_1 on y_1, t_2 on y_2*/

scalar psi=1 /*u on y_1, u on y_2*/
scalar chi=2 /*u on t_1*/
scalar gamma=3 /*u on t_2*/

scalar theta=0.5 /*t_1 on y_2*/
scalar kappa=0.3 /*t_2 on y_1*/
scalar tau=0.3 /*t_1 on t_2, t_2 on t_1*/

scalar omega=0.3 /*y_1 on t_2*/
scalar eta=0.3 /*y_1 on y_2*/
scalar lambda=0.3 /*y_2 on y_1*/

**generate variables

gen u=. /*unobserved fixed effect*/
gen t_1=. /*sibling 1's treatment*/
gen t_2=. /*sibling 2's treatment*/
gen y_1=. /*sibling 1's outcome*/
gen y_2=. /*sibling 2's outcome*/
gen d=. /*gain-score (y_2-y_1)*/

**simulation 

tempname m3c
postfile `m3c' sc_mc9 sc_ub_mc9 sc_lb_mc9 using m3c, replace
quietly {

forvalues i=1/$nmc {

	/*note: for y_1 and y_2 equations, "rnormal(0,1)" is residual*/

	replace u=rnormal(0,1)
	replace t_1=(chi*u)>0.5
	replace t_2=(gamma*u)>0.2
	replace y_2=(delta*t_2)+(theta*t_1)+(psi*u)+rnormal(0,1)
	replace y_1=(delta*t_1)+(lambda*y_2)+(psi*u)+rnormal(0,1)
	replace d=y_2-y_1
	
	/*gain-score regression*/
	reg d t_1 t_2
	
	/*lincom to compute spillover coefficient*/
	lincom t_1+t_2
	
	/*store lincom output*/
	scalar sc_mc9=r(estimate) /*spillover coefficient*/
	scalar sc_ub_mc9=(r(estimate)+(1.96*r(se))) /*95% CI upper bound*/
	scalar sc_lb_mc9=(r(estimate)-(1.96*r(se))) /*95% CI lower bound*/
	
	post `m3c' (sc_mc9) (sc_ub_mc9) (sc_lb_mc9)
}
}
postclose `m3c'

use m3c, clear
summarize

**store mean of results, save to append with other simulation results
egen sc=mean(sc_mc9)
egen ub=mean(sc_ub_mc9)
egen lb=mean(sc_lb_mc9)
gen id=9 /*ninth simulation*/

keep sc ub lb id
duplicates drop
duplicates report 

**save data 
save "sim9.dta", replace


/**********************/
*Graph Results
/**********************/

clear
set more off

**open first simulation dataset 
use "sim1.dta"

**append datasets (sim2-9)
append using "sim2.dta"
append using "sim3.dta"
append using "sim4.dta"
append using "sim5.dta"
append using "sim6.dta"
append using "sim7.dta"
append using "sim8.dta"
append using "sim9.dta"

**label id (assign to models)
label define model_lbl 1 "Figure 2.1A" 2 "1B" 3 "1C" ///
	4 "Figure 2.2A" 5 "2B" 6 "2C" ///
	7 "Figure 2.3A" 8 "3B" 9 "3C"
	
label values id model_lbl

**graph results

/*
NOTE: When importing Stata code into LaTeX, the compiler will not accept the en-dash in "{&theta} [MINUS] {&kappa}." Replace the hyphen with an en-dash to recreate the original figure.
*/

twoway(scatter id sc, mcolor(black) ysc(reverse) ///
		graphregion(color(white)) ///
		xtitle(Mean Spillover Coefficient (95% CI)) ///
		ytitle(Sibling Spillover Model) ///
		text(8.5 0.5 "({&theta} = 0.5)", size(small) /// 
			box margin(l+0.5 r+0.5 t+0.5 b+1)) /// 
		text(8.5 0.2 "({&theta} - {&kappa} = 0.2)", size(small) ///
			box margin(l+0.5 r+0.5 t+0.5 b+1)) ///
		xline(0.2, lcolor(black) lpattern(shortdash_dot)) ///
		xline(0.5, lcolor(black) lpattern(dash)) ///
		legend(off) xtick(#21)) ///
	(rcap ub lb id, horizontal lcolor(black)), ///
	ylabel(1(1)9, angle(horizontal) labsize(small) ///
		valuelabel notick) ///
	xlabel(-0.5(0.5)1.5, labsize(small)) saving(sim_results, replace)
	
\end{lstlisting}
\end{singlespace}

\end{document}